\documentclass[conference,compsoc]{IEEEtran}

\ifCLASSOPTIONcompsoc
  \usepackage[nocompress]{cite}
\else
  \usepackage{cite}
\fi

\ifCLASSINFOpdf
\else
\fi

\usepackage{makecell}
\usepackage{tikz}
\usepackage{amsmath,amsfonts}
\usepackage{algorithmic}
\usepackage{graphicx}
\usepackage{textcomp}
\usepackage{soul}
\usepackage{xcolor}
\usepackage{graphicx}
\usepackage{tabularx,booktabs}
\usepackage[table]{xcolor}
\usepackage{url}
\usepackage{color}
\usepackage{alltt}
\usepackage{syntax}
\usepackage{newfloat}
\usepackage{multicol}
\usepackage{amsmath}
\usepackage{tikz}
\usepackage{filecontents} % inlined bib file
\usepackage{graphicx}
\usepackage{epstopdf} %converting to PDF
\usepackage{xspace}
\usepackage{enumitem}
\usepackage{svg}
\usepackage{tabularx}
\usepackage[ruled,linesnumbered,noend]{algorithm2e}
\SetKw{Continue}{continue}
\usepackage{adjustbox}
\usepackage{wrapfig}
\usepackage{graphbox}
\usepackage{rotating}
\usepackage{pdfpages}
\usepackage{mdwlist}
\usepackage{multirow}
\usepackage{booktabs}
\usepackage{pgfplots}
\pgfplotsset{compat=1.18}
\usepackage{subcaption}
\usepackage{framed}
\usepackage{amsthm}

\usepackage{subcaption}
\usepackage[labelfont=bf,skip=0pt]{caption}
\usepackage{xurl}
\usepackage{hyperref}
\hypersetup{
	citebordercolor={0 1 0},
	citecolor=blue,
	colorlinks=false,
	filebordercolor={0 .5 .5},
	filecolor=green,
	linkbordercolor={1 0 0},
	linkcolor=blue,
	menubordercolor={1 0 0},
	pageanchor=true,
	pagebackref=true,
	pdfborder={0 0 1},
	pdfpagelabels=true,
	pdftex,
	plainpages=false,
	urlbordercolor={0 1 1},
	urlcolor=blue,
}

\usepackage[capitalize,nameinlink]{cleveref}
\hypersetup{%
	bookmarksnumbered, bookmarksopen=true, bookmarksopenlevel=1,%
}
\crefname{figure}{Figure}{Figures}
\crefname{listing}{Query}{Queries}
\crefname{section}{Section}{Sections}
\crefname{table}{Table}{Tables}
\crefname{BNF}{Grammar}{Grammars}
\crefname{algorithm}{Algorithm}{Algorithms}
\crefname{equation}{Equation}{Equations}

\usepackage{tcolorbox}
\tcbset{
  colback=gray!8,
  colframe=black,
  boxrule=0.5pt,
  arc=2pt,
  left=4pt,right=4pt,top=4pt,bottom=4pt
}
\usepackage{listings}
\definecolor{mygreen}{rgb}{0,0.6,0}
\definecolor{mygray}{rgb}{0.5,0.5,0.5}
\DeclareFloatingEnvironment[
fileext   = logr,% don't use log here!
listname  = {List of Grammars},
name      = Grammar,
placement = !htbp
]{BNF}

\newcommand{\msim}{\raise.17ex\hbox{$\scriptstyle\sim$}}

\newcommand{\eat}[1]{}

\newcommand{\sys}{\textsc{EARS}\xspace}  %**E**arly **A**nti-scam **R**ecognition **S**ystem

\usepackage{pifont}
\usepackage{amsmath}
\usepackage{pgfplots}

\newcommand{\distance}{8pt}
\usepackage{tikz}

\def\BibTeX{{\rm B\kern-.05em{\sc i\kern-.025em b}\kern-.08em
    T\kern-.1667em\lower.7ex\hbox{E}\kern-.125emX}}
\begin{document}
%-------------------------------------------------------------------------------

%don't want date printed
% \date{}

% \hyphenation{op-tical net-works semi-conduc-tor}

% \title{Demystifying Scam Playbooks:  Psychological Techniques, Scenario Taxonomy,  and Conversation-Aware Detection}

\title{The Anatomy of Scam Scenarios:  Large-Scale Characterization and Conversation-Aware Detection}
% Under the Hood of Online Scams:  Large-Scale Characterization and Conversation-Aware Detection
% \renewcommand\footnoterule{}

% Candidates
% The Anatomy of Scam Scenarios: Large-Scale Characterization and Conversation-Aware Detection
% From Scam Reports to Customer Conversations: Understanding and Detecting Scam Scenarios
% Demystifying Scam Scenarios: Large-Scale Characterization and Conversation-Aware Detection

%for single author (just remove % characters)
\author{
    Shang Ma$^\mathparagraph$, 
    Chen Yanai$^\ddagger$, 
    Avichai Ben$^\ddagger$, 
    Zichen Liu$^\dagger$,
    Yanfang Ye$^\mathparagraph$,
    Xusheng Xiao$^\dagger$
    \\

    \textsuperscript{$\dagger$}Arizona State University,
    \textsuperscript{$\ddagger$}Charm Security,    
    \textsuperscript{$\mathparagraph$}University of Notre Dame
    
    \\
{\small
    \texttt{avichai.ben@charmsecurity.com},
    \texttt{\{zliu396,xusheng.xiao\}@asu.edu},
    \texttt{\{sma5,yye7\}@nd.edu}    
}
      
}

\maketitle

\begin{abstract}
Online scams have become a pervasive global threat, causing substantial financial, psychological, and operational harm. Scammers embed psychological techniques (PTs) within reusable operational schemes to scale scam campaigns with minimal adaptation. However, existing studies often analyze PTs as isolated features, overlooking the recurring scam scenarios in which they are systematically deployed. To address this gap, we first conduct a large-scale empirical study to jointly characterize scam scenarios and their associated PTs. Specifically, we develop a data-driven pipeline to derive a hierarchical taxonomy of scam scenarios, consisting of 18 fine-grained scenarios grouped into 6 high-level tactics based on their PT profiles. Furthermore, to transfer this scenario-level knowledge to practical defense, we design a conversation-aware scam scenario detection approach for financial-institution customer interactions, enabling timely warning and intervention.

Our study on 102,054 real-world scam incident reports, spanning 2024-02-01 to 2025-10-31, reveals that PT usage is significantly associated with scam scenarios. We further show that scammers organize scenarios around different operational goals, such as broad victim exposure, high victim conversion, and high-value extraction, and reuse infrastructure, including IP addresses, domains, email addresses, and phone numbers, to launch coordinated campaigns at scale. 
Evaluation on 1,115 real-world customer-service conversations from an industry partner shows that our approach achieves 84.41\% tactic-level classification accuracy (84.14\% F1) and ranks the correct scam scenario within its top three predictions in up to 91.04\% of cases. These results demonstrate that scenario knowledge distilled from large-scale scam incident reports can effectively support early detection and characterization of scam-related customer-service conversations.
\end{abstract}

\section{Introduction}
\label{sec:intro}

% questions:why scenario, why customer-service conversation

% scam prevalent, severe, rising.

Online scams have become a global epidemic, inflicting severe financial and psychological harm on individuals and organizations. Their rapid proliferation is fueled by low-cost, fast, and often anonymous channels such as SMS and email. According to the U.S. Federal Trade Commission (FTC) 2025 data, consumers lost \$12.5 billion to fraud, a 25\% increase from the prior year, with investment fraud leading at \$5.7 billion in reported losses~\cite{ftcScamStats2025}. Identity theft cases exceeded 1.15 million, driven by sophisticated AI-assisted tactics including job offers, romance scams, and SMS-based fraud~\cite{ftcScam1,ftcScam2,ftcScam3}. Globally, Europe experienced billions in losses, with Germany and France alone reporting €10.6 billion and €7.6 billion~\cite{euroScam1,euroScam2,euroScam3}, while large-scale operations from East and Southeast Asia, such as crypto fraud and pig-butchering schemes, caused massive financial damage~\cite{asiaScam1,asiaScam2,asiaScam3}. AI-enabled impersonation scams accounted for over \$17 billion, and Southeast Asia’s scam ecosystem has evolved into a \$70B+ illicit marketplace often tied to forced-labor scam centers.

Beyond victim losses, scams impose significant operational, financial, and regulatory burdens on financial institutions. Banks must analyze vast volumes of communications (call transcripts, chat logs, emails) to detect potential fraud, often requiring manual review, inter-institution coordination, and evidence collection. Institutions also face liability for reimbursing victims of authorized push payment scams, where customers are tricked into transferring funds~\cite{psr2024app,ukfinance2024fraud}. Regulatory bodies have introduced stricter monitoring and reimbursement obligations, intensifying compliance pressure. These challenges underscore the urgent need for automated systems that identify scam signals in customer communications, enabling earlier intervention and reducing both investigative workload and financial losses.

The effectiveness and scalability of scams largely stem from the systematic use of psychological techniques (PT) and social engineering techniques, such as authority, urgency, reciprocity, and social proof, to manipulate victims’ decisions and actions~\cite{fernandes2024persuasion,longtchi2024quantifying}. These techniques are embedded within carefully crafted scam narratives that appear legitimate and contextually relevant~\cite{montanez2022cyber,longtchi2024internet}. For example, a smishing campaign may impersonate a highway toll service, claiming an unpaid \$12.51 fee that must be paid immediately via a provided link to avoid fines or license suspension~\cite{ic3toll,vakulov2025tollscam}. Such messages combine multiple PTs: \textit{authority} by impersonating official agencies, \textit{fear} through immediate penalties, and \textit{credibility} via links resembling official sites~\cite{cyberdefensetoll}. 
Once such a narrative proves effective, scammers can reuse it across victims and geographic regions with minor contextual adaptations, allowing them to scale operations while minimizing the cost of crafting new scams~\cite{krombholz2015advanced}.

Although prior work has studied PTs for scam detection and generation~\cite{schmitt2024digitaldeception,montanez2022cyber,longtchi2024internet,shen2025warned,sun2026prescam}, these studies often treat PTs as isolated persuasive signals and pay less attention to the recurring narrative structures in which they are deployed. In practice, scammers do not simply stack PTs arbitrarily. They embed them into reusable operational schemes, such as unpaid toll notifications, fake job offers, investment opportunities, or account-security alerts. These schemes define the scam’s premise, victim role, communication flow, and monetization path, while PTs shape how the victim is persuaded within that scheme.
Our preliminary study shows that scam narratives naturally group into recurring scam schemes, each characterized by a distinct combination of PTs. We refer to such a recurring operational scheme as a \textit{scam scenario}. 

To better understand scam scenarios and inform the development of effective defenses against online scams, we first conduct a large-scale empirical study based on real-world scam incident reports. We adopt a data-driven pipeline to discover a hierarchical taxonomy of scam scenarios, where fine-grained scenarios are organized into higher-level tactics based on their associations with PTs. We then conduct a detailed analytical study to examine the distribution of these scenarios, assess differences in their risk profiles, and characterize how scammers adapt their tactics and operations across scenarios. Building on the taxonomy and these empirical insights, we further develop an approach for detecting scam scenarios in customer-service conversations within financial institutions, where early recognition of the underlying scam scenario can support timely intervention~\cite{an2025revisiting,chehbouni2025enhancing}.

\noindent\textbf{Empirical Study of Scam Scenarios.}
Existing studies are often limited to small datasets, specific scam scenarios, or narrow time frames~\cite{ma2025psyscam,park2014scambaiter,miramirkhani2017dial,acharya2024conning,acharya2024explorative}. Consequently, there is still no systematic, large-scale analysis that jointly characterizes scam scenarios and the PTs used by scammers across scenarios. 
To bridge this gap, we curate a dataset of 102,054 scam incident reports from BBB Scam Tracker~\cite{bbbscamtracker}, covering the period from 2024-02-01 to 2025-10-31 (see~\autoref{fig:scamreport} for an example scam incident report).
We first leverage topic modeling to cluster scam reports based on recurring textual patterns, then apply LLM-assisted summarization, manual inspection, and expert-aligned refinement to consolidate noisy topics into 18 distinct scam scenarios. 
We then analyze PT usage within each scenario and find a significant association between scam scenarios and PTs ($\chi^2=28052$, $p<0.001$) revealing that PTs are not uniformly distributed across scenarios but instead reflect scenario-specific persuasion strategies. Based on this observation, we further organize the 18 scenarios into 6 higher-level tactics according to their shared PT profiles, as shown in \autoref{tab:scamtaxonomy}. 

Our analysis reveals several critical findings that fundamentally advance the understanding of how scammers structure, adapt, and scale their operations across scenarios. First, scammer weaponize PTs in scenarios by stacking them to exploit victims: 75.55\% of scam incidents contain at least two PTs, and each additional PT is associated with a 26.27\% increase in expected monetary loss. Second, different scenarios optimize for different operational goals, such as broad victim exposure, high victim conversion, or high-value extraction. Third, infrastructure analysis shows that scammers reuse operational resources to launch coordinated campaigns at scale: over 30\% of scam incidents are linked to distinct scam campaigns, and the largest campaign alone contains 3,809 incidents, spans all 18 scenarios. Finally, our case studies reveal that scammers adapt scenarios to seasonal events, cause harms beyond direct monetary loss, and operate international, cross-platform campaigns.

\definecolor{custclr}{HTML}{1F4E79}  % customer: deep blue
\definecolor{agntclr}{HTML}{9A2A2A}  % agent: deep red
\definecolor{custclr}{HTML}{1F4E79}  % customer: deep blue
\definecolor{agntclr}{HTML}{9A2A2A}  % agent: deep red

\newcommand{\spkC}[1]{\textcolor{custclr}{\textbf{c\textsubscript{#1}}}}
\newcommand{\spkA}[1]{\textcolor{agntclr}{\textbf{a\textsubscript{#1}}}}

\begin{figure*}[t!]
\centering
\footnotesize
\begin{tcolorbox}[
  colback=gray!3, colframe=black!55, boxrule=0.4pt, arc=2pt,
  left=6pt, right=6pt, top=5pt, bottom=5pt,
  fonttitle=\bfseries, coltitle=white, colbacktitle=black!55
]
\setlength{\parskip}{2pt}
\begin{multicols}{2}
\raggedright
\spkC{1}: Hello?\\
\spkA{1}: Hello, my name is David Okonkwo, and I'm calling from Bank of America Fraud Prevention. I'm calling regarding a recent transaction on your account that our system flagged as potentially unusual. Could I please confirm your full name and the last four digits of your account number?

\spkC{2}: Yes, this is Bradley Myers. My account number ends in 5486.\\
\spkA{2}: Thank you. We're looking at a payment of \$500 that was made recently. Do you recall making a \$500 payment using a gift card for a service?

\spkC{3}: Yes, I did make a payment of about \$500 recently, and it was with gift cards. It was for my utility bill, which was quite urgent.\\
\spkA{3}: We flagged this because using gift cards to pay a utility bill is quite unusual. Could you tell me why you used gift cards for this payment instead of a more traditional method?

\spkC{4}: The utility service said my power would be cut off very soon if I didn't pay immediately. They insisted on gift cards for a ``rapid processing payment'' to avoid the shutoff.\\
\spkA{4}: Which utility service did you speak with, and how did they instruct you to make this payment?

\spkC{5}: It was Idaho Power. They said it was the only way to process the payment quickly to avoid immediate power disconnection.\\
\spkA{5}: Legitimate utility companies, including Idaho Power, would never ask for payment using gift cards. They typically accept payments through their official website, by mail, or over the phone with a credit/debit card or bank transfer.

\spkC{6}: Oh dear. So, if they don't ask for gift cards, then\ldots was that not really Idaho Power? I was told it was the only way to prevent my power from being cut off immediately.\\
\spkA{6}: Based on what you've described, it sounds like you may have been targeted by an authority imposter scam. \textit{Scammers often pretend to be public agencies such as utility companies and demand immediate payment via gift cards to create a sense of urgency}.

\spkC{7}: It was a scam, then. I\ldots\ I did have a feeling, a slight unease, but they sounded so terribly official, and quite urgent. What does this mean for my account, exactly?\\
\spkA{7}: I'm very sorry to hear that. The \$500 payment you made with gift cards was likely sent to a scammer. We will mark this transaction as fraudulent and proceed with a fraud investigation, including steps to secure your account and explore recovery options.

\vspace{2pt}
\textit{\small [\,Agent initiates transfer to the specialized fraud department; call closes with a commitment to a callback within 5--10 minutes.\,]}
\end{multicols}
\end{tcolorbox}
\vspace{-4pt}
\caption{Example customer-service conversation in which the banking agent (\textbf{a}) investigates a flagged \$500 payment with the customer (\textbf{c}). In this case, the scammer impersonates the public utility service and threatens a power shutoff to pressure the victim into paying with gift cards. }
\label{fig:exampleConversation}
\end{figure*}

\noindent\textbf{Conversation-Aware Scam Scenario Detection}.
The preceding analysis studies scam scenarios from complete, retrospective reports. In practice, however, scam defense often happens during partial and unfolding interactions. In particular, when a suspicious transaction is flagged, the defender contacts the customer and asks about the incident. The customer then reveals information incrementally, such as how the interaction began, who the scammer claimed to be, and what payment was requested. As illustrated in \autoref{fig:exampleConversation}, the defender must infer the underlying scam scenario from incomplete and evolving customer utterances to enable timely intervention~\cite{an2025revisiting,chehbouni2025enhancing}. This setting motivates a key question: can scenario-level knowledge learned from large-scale scam reports support early scam scenario detection in customer-service conversations?

This problem can be formalized as a sequential text classification task, where the model updates its prediction as more conversation turns become available~\cite{serban2016hierarchical,raheja2019dialogue,qu2019intent}. However, applying this formulation to our setting faces three challenges. 
\textit{(1) Incomplete Context.} Customers typically describe incidents from memory, resulting in partial, fragmented, or uncertain context.  %To mitigate this challenge, we first pretrain the scenario classifier on a large corpus of scam incident reports from our empirical study, and then fine-tune it on customer-service conversation data. 
\textit{(2) Uncertain Evidence Onset.} The evidence needed to identify the underlying scam scenario may emerge only after several turns of a customer-service conversation. Early turns are often noisy, containing irrelevant information, such as greetings or identity verification.
\textit{(3) Diverse Narratives.} Real-world customer narratives vary widely in communication style, emotional state, and level of detail, even when they describe the same scam scenario. 
%To improve robustness,  we adopt an LLM-based data augmentation strategy that rewrites each training conversation with varied personas, narrative structures, opening styles, and lexical registers while preserving the underlying scenario narrative.
% \textit{(3) Long Context Windows.} Conversations often grow lengthy as the investigation progresses, exceeding the classification model's effective context window and introducing noisy turns. We therefore introduce a selective turn-dropping strategy that removes low-confidence classification turns to  preserve the turns most relevant to the underlying scenario.

%To remain uncommitted until sufficient evidence is observed while still leveraging earlier context, we retain the full conversation prefix and condition each turn on one of three progressive discourse phases: \textsc{Unrelated}, \textsc{Narrative}, and \textsc{Scam\_Evidence}.

Building on these insights, we present \sys (\textbf{E}arly \textbf{A}nti-scam \textbf{R}ecognition \textbf{S}ystem), which detects the underlying scam scenario from unfolding customer-service conversations. To handle incomplete context, \sys first pretrains a scenario classifier on the large-scale incident-report corpus. Although incident reports differ from conversations in format, they describe the same underlying scam scenarios from the victim's perspective, enabling the model to learn scenario-level semantics before adapting to conversation data. To handle uncertain evidence onset, \sys then finetunes the classifier on customer-service conversations under turn-level supervision: at each turn, it predicts from customer narrative observed so far and conditions the prediction on the turn's discourse phase, withholding judgment while the exchange is benign and committing to a scenario as soon as diagnostic evidence appears. To stay robust across the stylistic and structural variation of real customer narratives, \sys further augments the training conversations with an LLM-based, label-preserving rewriting pipeline. Together, these components turn report-level scenario knowledge into an early, deployable detector for live conversations.

We evaluate \sys on a test set of $1{,}115$ customer-service conversations comprising $14{,}406$ turns and covering all tactics and scenarios in our taxonomy. \sys achieves up to $84.41\%$ tactic-level accuracy ($84.14$ F1) and ranks the correct scam scenario among its top-3 predictions in over $90\%$ of conversation turns. Relative to a strong incident-report-pretrained encoder, our two conversation-level adaptations, LLM-based data augmentation and discourse-phase injection, jointly improve scenario macro-F1 from $64.87$ to $71.39$. \sys also outperforms LLM-based scam scenario detection: in scenario macro-F1, it exceeds the few-shot GPT-5.4 baseline, one of the strongest commercial LLMs, by $25.62$ points.
These results show that the scenario taxonomy derived from real-world incident reports transfers to customer-service conversations and supports earlier recognition of scam-related customer cases.

% Our contributions are as follows:
% \begin{itemize}[noitemsep, topsep=1pt, partopsep=1pt, listparindent=\parindent, leftmargin=*]
% \item We conduct a large-scale empirical study of scam scenarios using 102,054 real-world scam incident reports and derive a hierarchical taxonomy of 18 scenarios grouped into 6 higher-level tactics.
% \item We uncover empirical findings about the scam ecosystem and scammer strategies, including scenario PT association, cross-scenario operation tactics, and infrastructure abuse in coordinated campaigns.
% \item We develop e a scam scenario detection approach for customer-service conversations in financial institutions and evaluated on [xxx].
% \end{itemize}

% imcomplete context, diverse noise, long context window

\section{Background}

\subsection{Human Vulnerabilities and Psychological Exploits}
Human vulnerabilities are intrinsic cognitive, emotional, and social tendencies that influence how people perceive risk, trust others, and make decisions~\cite{sharma2020analysis}. Human-centric cyber attacks exploit these vulnerabilities through psychological rather than purely technical means~\cite{montanez2022cyber,longtchi2024internet}. Prior work has characterized these manipulation strategies as \emph{Psychological Techniques} (PTs)~\cite{ma2025psyscam} and Human Vulnerability Exploits (HVEs)~\cite{charmhve}, both describing recurring methods used by scammers to influence victims. Building on these frameworks, we refine and extend the PT taxonomy using the HVE framework, yielding nine PT categories: \textit{Authority}, \textit{Phantom Riches}, \textit{Fear and Intimidation}, \textit{Liking}, \textit{Urgency and Scarcity}, \textit{Pretext and Trust}, \textit{Evoking Social Norms}, \textit{Consistency}, and \textit{Social Proof}. Definitions are provided in Appendix~\autoref{tab:ptdefinitions}.

\subsection{Classification of Scam Types}
A scam taxonomy is essential for understanding, measuring, and mitigating online scams, providing researchers with a common vocabulary for analysis and practitioners with a framework for education, triage, and defense~\cite{charmhve}. Existing resources such as the BBB Scam Tracker glossary define 32 scam types~\cite{bbbscamtype}, but these labels are often flat, coarse, and semantically overlapping; for example, the category \textit{phishing} may encompass distinct scenarios such as delivery impersonation, account-security alerts, fake subscriptions, and credential-theft campaigns (\autoref{fig:scamreport}). More systematic efforts, such as the Stanford Center on Longevity and FINRA fraud taxonomy, organize fraud into hierarchical dimensions including claimed identity, victim-facing pretext, promised benefit or threatened loss, and requested victim action~\cite{beals2015framework}. However, manually curated taxonomies are costly to maintain and struggle to keep pace with evolving scam ecosystems, which increasingly incorporate new technologies, platforms, and monetization models such as cryptocurrency scams, data-breach extortion, and remote-work employment fraud. These challenges motivate a data-driven, continuously updateable taxonomy derived from real-world scam reports.

% In this work, we construct of our taxonomy is based on the methodology of this.

\section{Empirical Study of Scam Scenarios}

\begin{figure}[t]
\centering
\footnotesize
\begin{tcolorbox}[
  colback=gray!3, colframe=black!55, boxrule=0.4pt, arc=2pt,
  left=5pt, right=5pt, top=4pt, bottom=4pt,
  title=\small\textbf{\textcolor{white}{\nolinkurl{https://www.bbb.org/scamtracker/lookupscam/1291795}}},
  fonttitle=\bfseries, coltitle=white, colbacktitle=black!55
]
\textbf{Description.} I was on the phone with the USPS and the USPS
online trying to locate a package and it came up to fill out a form
to be able to track a package, it then asks for a \$5 refundable
deposit, USPS advised me that was not them. I called and assumed
this was resolved. A charge was taken out of my account for \$65 for
a service I did not ask for. When I called to see, it was not even
on my phone number, but a work cell number, they said they cx the
``subscription'' I didn't know I had but would not refund my money.
\vspace{3pt}\hrule\vspace{3pt}
{\setlength{\tabcolsep}{3pt}\renewcommand{\arraystretch}{1.05}%
\begin{tabular}{@{}p{0.47\columnwidth}p{0.47\columnwidth}@{}}
\textbf{Scam Type:} Phishing      & \textbf{Scam ID:} 1291795        \\
\textbf{Scammer:} (800) 556-3410 & \textbf{Victim Loc.:} OK,  73065 \\
\textbf{Date:} May 15, 2026       & \textbf{Dollars Lost:} \$65       \\
\textbf{Email:} info@justanswer.com & \textbf{URL:} justanswer        \\
\end{tabular}}
\end{tcolorbox}
\vspace{-4pt}
    \caption{Example of a scam incident report.}
  \label{fig:scamreport}
\end{figure}

% \noindent\textbf{Challenge.} 
% As shown in \autoref{fig:scamreport}, BBB provides an official
% \texttt{Scam Type} label for each report describing the scenario. However, these labels are often
% too coarse, inaccurate, or semantically overlapping. For example, the
% label \textit{phishing} is highly ambiguous and may include many
% substantively different scam scenarios. 
% Therefore, rather than relying
% on the platform-provided labels, we adopt a data-driven pipeline to discover
% a taxonomy of scam scenarios from the large corpus of scam incident reports. We
% then annotate the dataset based on this taxonomy and train a supervised
% deep-learning model to automatically classify the scam scenario of each
% report.

% \subsection{Study Overview}
% \label{subsec:problemOverviewScenario}
In our empirical study, we first employ a data-driven pipeline to extract scam scenarios from a large corpus of real-world scam incident reports. Building on these extracted scenarios, we next examine their relationships with PTs and systematically characterize them along multiple dimensions, including risk profiles, scammers’ operational strategies, and infrastructure usage.
% To formalize this notion, we define a scam scenario as follows:

% \begin{definition}[Scam Scenario]
% A scam scenario is a recurring operational scheme that captures the core strategy used by a scammer, including the claimed pretext, the victim-facing narrative, and the requested action or payment method.
% \end{definition}
% For example, a utility shutoff scam, a fake job offer
% scam, and a bank account alert scam correspond to different scenarios
% because they rely on different pretexts, communication flows, and victim
% actions.

\subsection{Datasets}
We collect scam incident reports from BBB Scam Tracker~\cite{bbbscamtracker}, a public platform where victims and targets of scams voluntarily submit reports containing rich textual descriptions and structured metadata. Starting from 177,989 raw reports, we apply a two-stage preprocessing pipeline. 
First, we remove business-fraud cases outside the scope of this study (e.g., undelivered online purchases) using a lightweight LLM-based classifier (Qwen3-1.7B); manual evaluation of 100 sampled reports found no misclassified irrelevant cases.
Second, we remove duplicate submissions and filter out overly short reports ($\leq$27 words, corresponding to the 20th percentile in length) that are unlikely to contain useful information. After preprocessing, the final dataset contains 102,054 high-quality scam incident reports spanning 2024-02-01 to 2025-10-31.

% \autoref{tab:} shows the overall statistic of our dataset after preprocessing.

% \subsection{Dataset Analysis}

% % phone
% % ip,domain
% % email
% % location, time
% % scam type

% In this subsection, we present the data analysis on the metadata of the collected scam incident reports.

\subsection{Data-Driven Scam Scenario Discovery}
\label{subsec:topicModeling}
%finding: scam sharing scenario
% cluster figure, cluster error to prove clustering is good

\noindent\textbf{Topic Modeling.}
Scammers often reuse effective scenarios across victims and geographic regions with only minor contextual adaptations. To capture such recurring patterns, we apply topic modeling to partition a large corpus of scam incident reports into clusters, where each cluster serves as a candidate scam scenario.
Specifically, we employ BERTopic~\cite{grootendorst2022bertopic}, which first embeds each report into a dense semantic space using a pretrained language model, then clusters semantically similar reports into preliminary topics, and finally extracts representative topic keywords via class-based TF-IDF. We set the number of topics to 100, intentionally overestimating the expected number of true scam scenarios. This over-segmentation reduces cluster heterogeneity and more effectively isolates outliers, i.e., reports that do not clearly align with any coherent topic.
Overall, the model assigns 64,963 of 102,054 reports to 99 non-outlier topics, while the remaining 36.34\% are marked as outliers. This relatively high outlier rate is expected, given the noisy and heterogeneous nature of real-world scam incident reports, and it helps prevent ambiguous cases from being forced into poorly formed clusters. \autoref{tab:bertopictopicssummary} in Appendix gives examples of the BERTopic outputs. 

The clustering metrics show that the raw topics are locally coherent but globally overlapping. 
Specifically, the average centroid similarity is 0.6769, indicating that reports within the same topic are semantically close, and the keyword diversity score is 0.8677, suggesting that each topic captures distinct recurring narratives.  However, the cosine silhouette score is only 0.0395, meaning that many topics are weakly separated from nearby topics in the embedding space. 
Thus, the predefined 100-topic setting yields fine-grained clusters that often split the same scam scenario into multiple semantically related topics. To consolidate these raw topic clusters into a smaller set of distinct scam scenarios, we apply a two-step LLM-human collaborative annotation pipeline.

\noindent\textbf{Topic Merging and Scenario Derivation.}
Based on the top-50 keywords associated with each topic, we prompt an LLM to generate a concise one- to three-sentence summary (see prompt template in \autoref{fig:topicSummaryPrompt}), with representative examples provided in \autoref{tab:bertopictopicssummary}. BERTopic further provides a relevance score between each report and its assigned topic, enabling identification of the most representative reports per topic. We then perform a manual inspection of topic keywords, LLM-generated summaries, and the top-5 representative reports, and merge topics that correspond to the same underlying scam scenario but differ due to lexical variations or reporting styles. This consolidation reduces the initial 99 topics to 18 distinct scam scenarios.
At the same time, we create an initial name and definition of each scenario during this annotation process, and align them with feedback from security experts at Charm Security and existing scam classification systems~\cite{beals2015framework,bbbscamtype}. This step ensures that each scenario is semantically accurate, distinct from the others, grounded in the data, and consistent with practitioner-facing terminology.  

\noindent\textbf{Scenario Refinement.}
Topic modeling provides only rough scenario clusters. For example, a scam incident report may be assigned to an incorrect cluster because it shares similar wording with reports from another scenario. Thus, topic modeling alone is insufficient for producing high-quality scenario labels for subsequent analysis. However, manually annotating the entire dataset is expensive. To address this, we manually annotate reports from each scenario cluster, and refine the definition of each scenario. This process is part of our active learning pipeline, which will be elaborated in Subsection~\ref{subsec:incidentPretrain}.

\begin{table*}[t]
\centering
\setlength{\tabcolsep}{1.3pt}
\renewcommand{\arraystretch}{1.2}
\footnotesize
\caption{Hierarchical taxonomy of scam scenarios grouped by major psychological technique drives}
\begin{tabular}{p{2.8cm} >{\scriptsize}p{9.1cm} p{5.4cm}}
\toprule
\textbf{Scenario} & \textbf{Description} & \textbf{Tactic} \\
\midrule
Retail & Impersonating a retail brand, selling counterfeit goods, or fake ads for non-existent deals. & \multirow{5}{=}[0pt]{\raggedright\textbf{Consumer \& Service Scams}\par\textit{\scriptsize Scams that exploit trust in legitimate businesses, brands, or service providers by impersonating them or offering fraudulent products/services.}\par\smallskip PT Drive: Credibility} \\
Insurance \& Warranty & Impersonating a provider, alleging policy issues, or offering fake health/auto/home policies. & \\
E-Commerce & Impersonating e-commerce platforms or delivery services, or purchase scams where items are never delivered. & \\
Financial Services & Impersonating a bank/credit card company, alleging account problems, or offering fraudulent debt relief. & \\
Tech \& Online Service & Impersonating a tech company, alleging subscription/virus issues, fake tech support, or recovery scams. & \\
\midrule
Government & Impersonating the government, such as the IRS, SSA, or other agencies to demand fake payments or steal personal information. & \multirow{2}{=}[0pt]{\raggedright\textbf{Authority \& Compliance Scams}\par\textit{\scriptsize Scams that leverage the perceived authority to coerce victims through fear and intimidation.}\par\smallskip PT Drive: Authority, Fear and Intimidation} \\
Legal & Impersonating the law enforcer such as court, lawyer, or the FBI to allege warrants or criminal involvement, demanding payment to ``resolve'' it. & \\
\midrule
Lottery \& Sweepstakes & Victims are offered lottery, prize, or products (e.g., donations, heritage, fake social media giveaways) but must pay a ``fee'' or ``tax'' to claim winnings. & \multirow{4}{=}[0pt]{\raggedright\textbf{Windfall Scams}\par\textit{\scriptsize Scams that lure victims with promises of unexpected financial gains, prizes, or exceptional deals.}\par\smallskip PT Drive: Phantom Riches} \\
Funds, Grants \& Aid & Fake grants, scholarships, or aid programs requiring upfront fees or personal information. & \\
Investment \& Trading & Promising high-return, low-risk investments via fraudulent platforms, fake stock tips, etc. & \\
Good Deals & Fake check/overpayment scams, fraudulent travel packages, or ads for non-existent deals. & \\
\midrule
Sextortion & Threatening to have compromising photos or videos of the victim and threatens to release them unless a payment (often crypto) is made. & \multirow{2}{=}[0pt]{\raggedright\textbf{Extortion Scams}\par\textit{\scriptsize Scams that coerce victims into payment via extortions.}\par\smallskip PT Drive: Fear and Intimidation} \\
Hack \& Data Breach & Threatening to have hacked the victim's device, encrypted their files (ransomware), or stolen their data, demanding payment for its return. & \\
\midrule
Fake Job Offer & Fake applications to steal personally identifiable information, or requiring payment for ``training''/``equipment'' for non-existent jobs. & \multirow{2}{=}[0pt]{\raggedright\textbf{Employment Scams}\par\textit{\scriptsize Scams that exploit job seekers by offering fake employment.}\par\smallskip PT Drive: Phantom Riches, Consistency} \\
Unpaid Labor & Tricking victims into performing ``trial'' work with a promise of payment that never comes, such as package delivery, data labeling. & \\
\midrule
Pig Butchering & Building a fake relationship to manipulate the victim into sending money or investing in fraudulent platforms. & \multirow{3}{=}[0pt]{\raggedright\textbf{Relationship \& Trust Scams}\par\textit{\scriptsize Scams that build or exploit personal and emotional connections to manipulate victims}\par\smallskip PT Drive: Evoking Social Norms} \\
Charity & Soliciting donations for fake charitable causes, such as disasters, wars. & \\
Friends \& Relatives & Claiming to be a friend/relative member and needing financial help due to an emergency. & \\
\bottomrule
\end{tabular}
\label{tab:scamtaxonomy}
\end{table*}

\subsection{Taxonomy of Scam Scenarios}
\label{subsec:scamtaxonomy}
In total, our scenario discovery process yields 32,746 scam incident reports annotated with 18 scam scenarios, with their distribution reported in Appendix~\autoref{fig:scenarioDistribution}. To further understand how different scenarios exploit victims, we next analyze their associated PTs. Specifically, we build a PT classifier following prior work~\cite{ma2025psyscam}, apply it to our dataset, and examine the prevalence of PTs across scenarios.

\begin{figure}[t]
    \centering
    \includegraphics[width=\columnwidth]{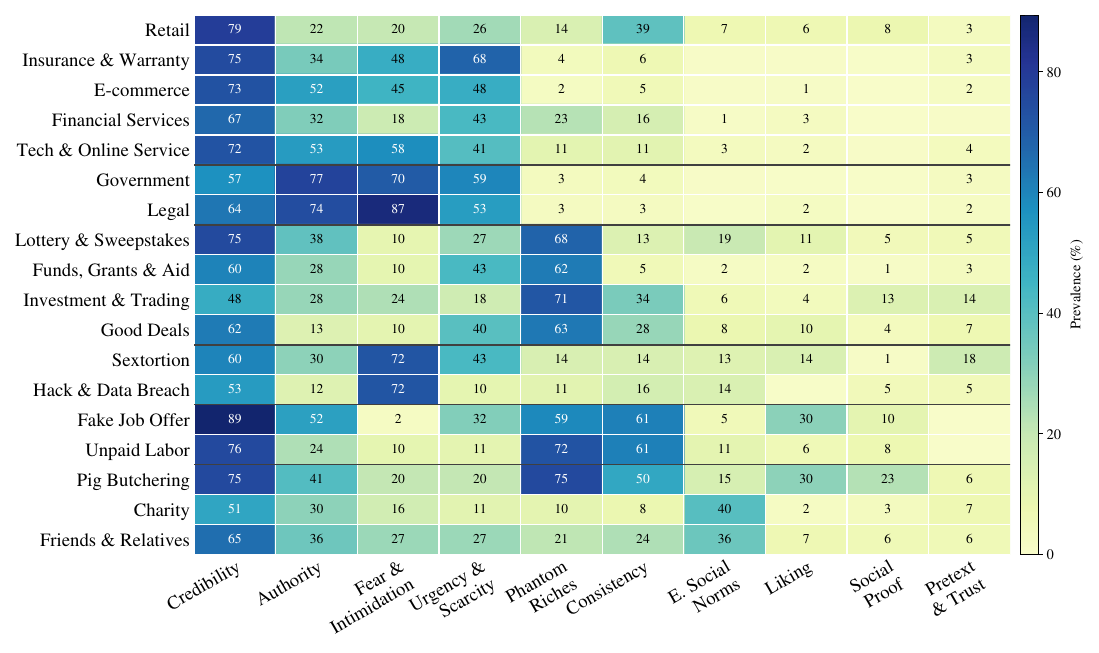}
    \caption{PT distribution across scam scenarios (Finding~1).}
    \label{fig:ptPrevalenceScenario}
\end{figure}

\noindent\textbf{Finding 1: }\textit{Scam scenarios exhibit distinct but recurring psychological technique profiles.}
Our chi-square test shows a significant association between scam scenarios and PTs
($\chi^2=28052$, $df=153$, $p<0.001$), with a moderate effect size measured by
Cramer's $V=0.195$. This indicates that PTs are not uniformly distributed across scenarios. As shown in \autoref{fig:ptPrevalenceScenario}, \textit{Credibility} is widely used across nearly all scenarios, which is intuitive because most scams leverage impersonation to gain victims' trust. Beyond this common pattern, however, scenarios show distinctive PT profiles. For example, Government and Legal scams rely heavily on \textit{Authority} and \textit{Fear \& Intimidation}, while Charity and Friends \& Relatives scams show stronger use of \textit{Evoking Social Norms}. These patterns suggest that scammers craft different victim-facing narratives while exploiting similar underlying psychological levers.

To capture this tactic-level regularity, we introduce an additional categorization layer above the 18 scenarios. This layer groups scenarios that share similar PT drives and scammer tactics. The resulting hierarchical taxonomy is shown in \autoref{tab:scamtaxonomy}, where each high-level category groups tactically related scenarios under a major PT drive.

\subsection{Analysis}
As shown in \autoref{fig:scamreport}, each scam incident report contains rich telemetry about the scammer and the incident, enabling us to examine how these signals characterize different scam scenarios. In this part, we primarily analyze scenarios at the tactic level, as it combines scenario semantics with PTs and provides a cleaner basis for visualization. Fine-grained scenario-level results are provided in Appendix~\ref{appendix:pt}.

\begin{figure}[h]
    \centering
    \includegraphics[width=\columnwidth]{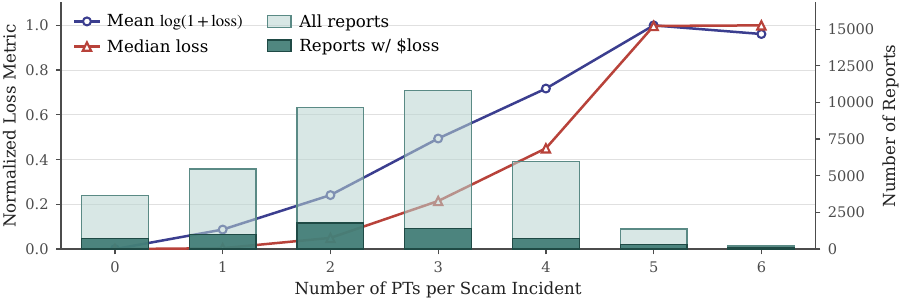}
    \caption{PT combinations and dollar lost  (Finding~2).}
    \label{fig:ptCountDistribution}
\end{figure}

% which scenario stack most, why scenario not with specific PT; why not as many PTs as possible; most universal PT
\noindent\textbf{Finding 2: }\textit{Scammers weaponize PTs by stacking them to manipulate victims: each additional PT coincides 26.27\% increase in expected monetary loss.}
As shown in \autoref{fig:ptCountDistribution}, PT stacking is common in real-world scams: $75.55\%$ of incident reports contain at least two PTs. Moreover, reports with more PTs tend to incur higher losses, suggesting that psychological complexity is associated with scam severity. However, effective PT stacking is scenario-dependent rather than simply additive. As shown in \autoref{fig:ptPrevalenceScenario}, \textit{Government} and \textit{Legal} scams rely heavily on \textit{Authority} and \textit{Fear and Intimidation} (both $>70\%$) while rarely using \textit{Liking}, \textit{Social Proof}, or \textit{Evoking Social Norms}, which are difficult to reconcile with threat-based narratives. These findings suggest that scammers combine multiple PTs only when they are coherent with the underlying scam playbook.

% In other words, scams that deploy multiple psychological levers appear more
% dangerous, as they can simultaneously build trust, create pressure, and motivate
% victims toward harmful actions.

We further validate this trend statistically. Using 6,050 reports with valid
loss values, we find a significant positive Spearman correlation between the
number of PTs and monetary loss ($\rho=0.398$, $p=1.07\times10^{-228}$). The
association remains significant after controlling for scam scenario and report
length using a regression model with heteroscedasticity-robust standard errors:
each additional PT is associated with a 26.27\% increase in expected monetary
loss ($p<0.001$, 95\% CI: 10.29\%--44.56\%). These results suggest that PT combinations are linked to scam severity, as multiple psychological levers can simultaneously build trust, create pressure, and drive harmful actions.

\begin{figure}[h]
    \centering
    \includegraphics[width=\columnwidth]{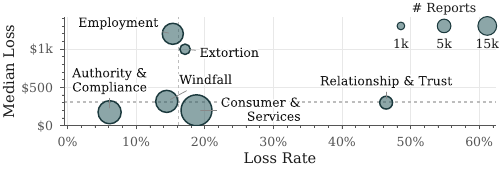}
    \caption{Incident report volume, monetary loss rate, and median loss across scenario tactics (Finding~3).}
    \label{fig:ScenarioPSC}
\end{figure}
\noindent\textbf{Finding 3: }\textit{Scammers tune scenarios and tactics for different strategic goals: mass exposure, high victim conversion, or high-value extraction.}
\autoref{fig:ScenarioPSC} characterizes each scam scenario tactic along three dimensions:
prevalence, measured by the number of reports; conversion rate, measured by the
fraction of reports with positive monetary loss; and severity, measured by
median dollar loss. We observe a diversity of tactic profiles among these dimensions. 
For example, Consumer \& Services and Authority \& Compliance scams account for the most incident reports, suggesting broad exposure but their median losses are relatively low, indicating limited per-victim financial harm. 
 In contrast, Relationship \& Trust scams have the
highest conversion rate, with nearly half of reports involving monetary loss,
consistent with their personalized nature: scammers either gradually build
emotional trust (Pig Butchering) or impersonate existing social relationships (Friends \& Relatives). Employment and
Extortion scams show a different profile: they are lower in volume and
conversion, but produce the highest typical losses. Employment scams often
exploit desperate and financially vulnerable job seekers who are willing to pay more fees for a potential job opportunity, while Extortion scams rely on threats or
ransom-like pressure that could lead to severe reputation harm (Sexortion) or personal loss (Hack \& Data Breach). Overall, these
results indicate that scammers optimize tactics with different goals: exposure, victim conversion, and profit gain.

\begin{figure}[h]
    \centering
    \includegraphics[width=\columnwidth]{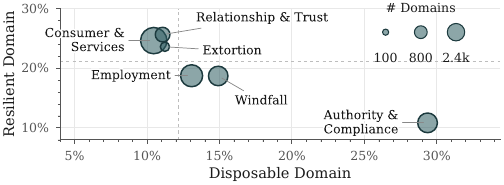}
    \caption{Resilient and disposable domain rates across tactics (Finding~4).}
    \label{fig:domainBubble}
\end{figure}

\noindent\textbf{Finding 4: }\textit{Scammers exploit infrastructure for different strategic goals across scenarios and tactics: disposable domains for mass exposure, and resilient domains for sustained engagement.}
We further analyze scam campaign infrastructure using website domains extracted from incident reports. Among 37,197 reports, 8,595 contain valid domains, yielding 5,486 unique domains. For each domain, we collect WHOIS, DNS, ASN/hosting, CDN/Cloudflare usage, and HTTP liveness signals, and classify domains as \emph{disposable} (e.g., newly registered, short-lived within one year, or unreachable) or \emph{resilient} (e.g., persistently live and CDN/Cloudflare-backed). 
As shown in \autoref{fig:domainBubble}, tactics exhibit distinct infrastructure patterns. \textit{Authority \& Compliance} scams have the highest disposable-domain rate and the lowest resilient-domain rate, reflecting rapid deployment and abandonment consistent with broad-exposure campaigns. In contrast, \textit{Relationship \& Trust} scams show the highest resilient-domain rate and lower disposable usage, consistent with sustained infrastructure needed for prolonged trust-building and repeated interactions. Overall, scam infrastructure is not uniform but adapted to the operational goals of each tactic.

\noindent\textbf{Finding 5: }\textit{Scammers reuse infrastructure to coordinate campaigns: over 30\% of scam incidents are linked to distinct campaigns, and the largest campaign alone accounts for more than 10\% of all incidents.}
To explore whether the network infrastructure is reused across incidents, we build a graph to connect scam incidents potentially operated by the same scammers. 
Specifically, each node is a scam incident, and two incidents are connected if they share an IP address, domain, email address, or phone number. 
Each connected component of the graph therefore represents a potential scam campaign. 
\autoref{tab:scamClusters} summarizes the five largest components by number of reports, total loss, shared infrastructure, and covered tactics. 
The largest component, $\mathcal{C}_1$, contains 3,809 reports, causes \$559K in reported losses, and spans all tactics and scenarios, accounting for more than 10\% of all incidents in our dataset. 
This data reveals complex cross-tactic infrastructures that allow scammers to scale and sustain their operations.

\begin{table}[t]
\centering
\footnotesize
\setlength{\tabcolsep}{1.5pt}
\renewcommand{\arraystretch}{1.15}
\caption{Top scam campaign clusters identified in our dataset. Infra. (D/I/E/P) denotes the number of shared domains, IPs, emails, and phone numbers within the cluster (Finding~5).}
\begin{tabular}{@{}lrrlrr@{}}
\toprule
\textbf{Cluster} & \textbf{\#Reports} & \textbf{Loss (\$)} & \textbf{Infra. (D/I/E/P)} & \textbf{\#Tactics} & \textbf{\#Scenarios} \\
\midrule
$\mathcal{C}_1$ & 3{,}809 & 559{,}666 & 494 / 209 / 665 / 1{,}116 & 6 & 18 \\
$\mathcal{C}_2$ &   698   &   6{,}010 &   0 /   0 /  16 /      21 & 5 & 12 \\
$\mathcal{C}_3$ &   350   &   1{,}643 &   1 /   0 /   7 /       2 & 4 & 13 \\
$\mathcal{C}_4$ &   307   &   6{,}176 &   5 /   4 /   5 /      11 & 6 & 9 \\
$\mathcal{C}_5$ &   287   &  44{,}084 & 211 /   7 / 183 /      78 & 5 & 6 \\
\bottomrule
\end{tabular}

\label{tab:scamClusters}
\end{table}

\subsection{Case studies}

\begin{figure}[h]
    \centering
    \begin{subfigure}{0.48\columnwidth}
        \centering
        \includegraphics[width=\linewidth]{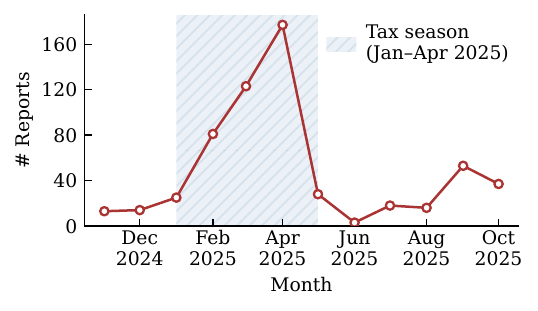}
        \caption{IRS impersonation}
        \label{fig:irsImpersonation}
    \end{subfigure}
    \hfill
    \begin{subfigure}{0.48\columnwidth}
        \centering
        \includegraphics[width=\linewidth]{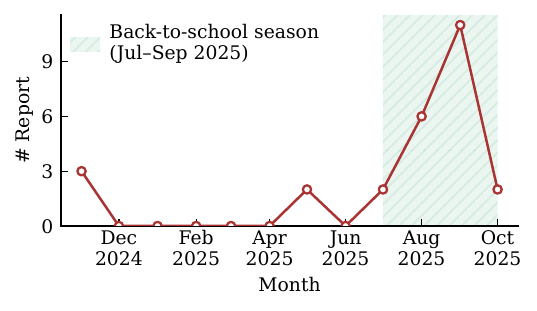}
        \caption{Student loan scams}
        \label{fig:btsLoanScam}
    \end{subfigure}
    \caption{Case studies on seasonal scam trends.}
    \label{fig:temporalCaseStudies}
\end{figure}

\noindent\textbf{Case Study 1: Seasonal scam events.}
We examine whether scam narratives align with real-world seasonal events, focusing on the U.S. tax season and the back-to-school season. For tax-related scams, we study IRS impersonation scams, which belong to the ``Authority \& Compliance--Government'' scenario; for back-to-school scams, we study student loan scams, which belong to the ``Windfall--Funds, Grants \& Aid'' scenario. We identify both cases using keywords over the \texttt{description} field of scam incident reports. As shown in \autoref{fig:temporalCaseStudies}, IRS impersonation scams rise sharply during Jan--Apr 2025 and peak around the tax filing deadline, while student loan scams increase during Jul--Sep 2025, aligning with the back-to-school period. These patterns suggest that scammers adapt their narratives to timely real-world contexts, exploiting periods when victims are more attentive to specific institutions, deadlines, or financial needs.

\begin{figure}[h]
  \centering
  \begin{subfigure}{0.48\columnwidth}
    \centering
    \includegraphics[width=\columnwidth]{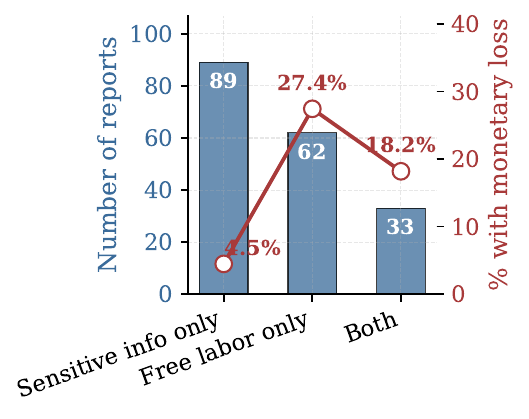}
    \caption{Different scam outcomes.}
    \label{fig:outcomeLoss}
  \end{subfigure}
  \hfill
  \begin{subfigure}{0.48\columnwidth}
    \centering
    \includegraphics[width=\columnwidth]{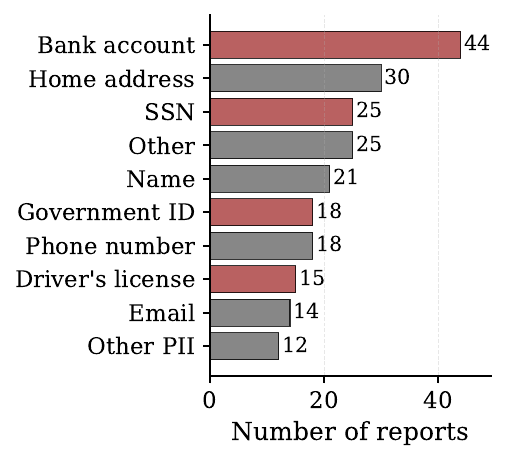}
    \caption{Sensitive information leak.}
    \label{fig:sensitiveInfo}
  \end{subfigure}
  \caption{Case studies on employment scams.}
  \label{fig:outcomes}
\end{figure}

\noindent\textbf{Case Study 2: Victims lose more than money.}
We manually annotated 500 randomly sampled Employment-scam incident reports and recorded two non-monetary outcomes: sensitive information disclosure (e.g., SSN, bank accounts, and government IDs) and unpaid labor (e.g., package reshipping, fake reviews, and mystery shopping). As shown in ~\autoref{fig:outcomes}, scammers frequently obtain identity- and finance-critical information, with bank accounts, home addresses, SSNs, and government IDs among the most common disclosures. These harms can expose victims to long-term identity theft. In contrast, free-labor scams are also more likely to involve monetary loss than sensitive-info-only scams (27.4\% vs.~4.5\%), because victims are often asked to pay upfront costs such as shipping, gift cards, or training fees.

% \autoref{tab:EmploymentChannelPattern}
% \begin{table}[h]
% \centering
% \small
% \setlength{\tabcolsep}{5pt}
% \renewcommand{\arraystretch}{1.1}
% \begin{tabular}{lrrr}
% \toprule
% \textbf{Contact Channel} & \textbf{\# Reports} & \textbf{Sens.\ Rate (\%)} & \textbf{Labor Rate (\%)} \\
% \midrule
% Private channel only           & 199 & 28.6 & 20.1 \\
% Unknown                        & 186 & 19.4 & 18.8 \\
% Platform $\rightarrow$ private &  67 & 37.3 & 23.9 \\
% Platform only                  &  20 & 20.0 & 20.0 \\
% \midrule
% \textbf{Total}                 & \textbf{472} & \textbf{25.8} & \textbf{20.6} \\
% \bottomrule
% \end{tabular}
% \end{table}
% \caption{Distribution of scam incident reports by communication channel pattern.}
% \label{tab:EmploymentChannelPattern}
% \end{table}

\begin{figure}[h]
    \centering
    \includegraphics[width=\columnwidth]{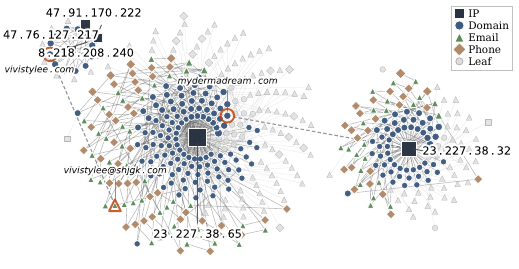}
    \caption{Visualization of the $C_5$ scam campaign cluster. Nodes with degree greater than one are colored by infrastructure type, while degree-one nodes are shown as gray leaves. The top-degree IP nodes and subcluster connection nodes are labeled.}
    \label{fig:c5Network}
\end{figure}

\noindent\textbf{Case Study 3: An international and cross-platform counterfeit scam.}
We conduct a case study on cluster $C_5$ in \autoref{tab:scamClusters}. As shown in \autoref{fig:c5Network}, the cluster is anchored by three dense subclusters: two Shopify-resolved IPs (23.227.38.65 and 23.227.38.32) and one Alibaba Cloud subcluster centered on three co-hosting IPs (47.76.127.217, 47.91.170.222, and 8.218.208.240). These subclusters are linked via a shared identity, where the domain \textit{vivistylee.com} (Alibaba Cloud subcluster) connects through the email \textit{vivistylee@shjgk.com} to storefronts in the Shopify subcluster. Manual inspection of incident reports further reveals a consistent cross-platform scam flow (see~\cite{sundanceFb,sundanceIns} for the corersponding scam incident reports): scammers register domains and emails (e.g., \textit{vivistylee.com}, \textit{vivistylee@shjgk.com}) on Alibaba Cloud infrastructure, then run Facebook and Instagram ads impersonating brands such as Sundance Catalogue to promote heavily discounted products. Victims who click these ads are redirected to seemingly legitimate storefronts, place orders, and ultimately receive suspicious or counterfeit shipments from China. Overall, this case illustrates how scammers integrate cloud infrastructure, e-commerce storefronts, reusable email identities, and social media advertising to execute an international counterfeit-shopping campaign.

% \noindent\textbf{Active Learning.}

\begin{figure*}[t]
  \centering
  \includegraphics[width=\linewidth]{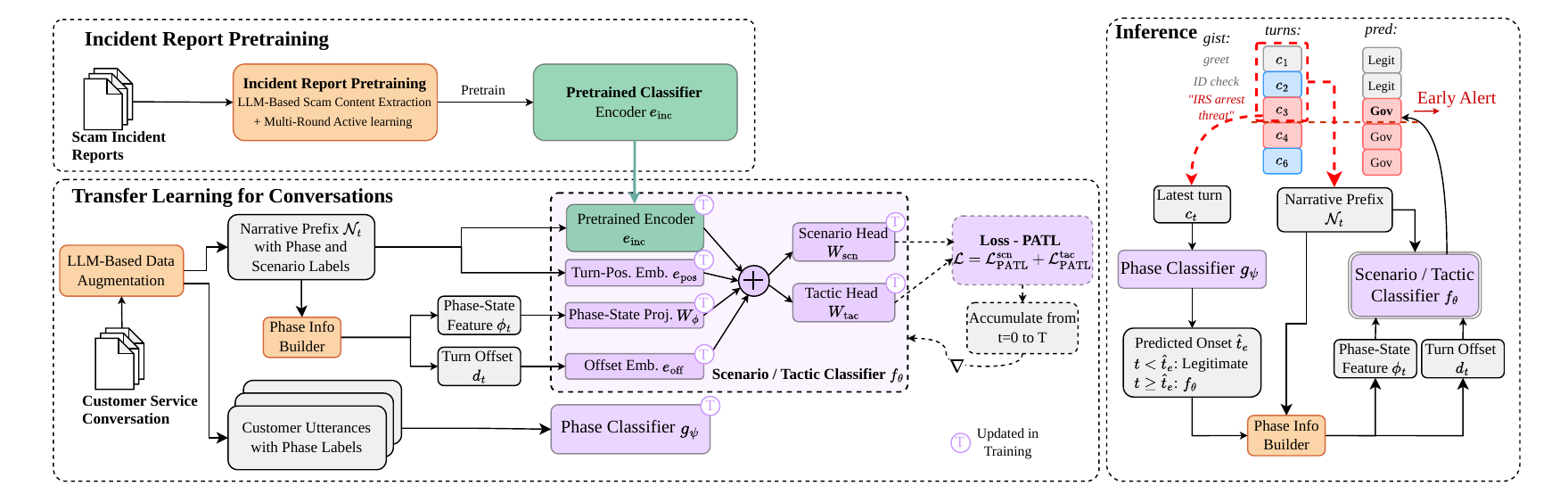}
  \caption{Overview of \sys.}
  \label{fig:methodoverview}
\end{figure*}

\section{Conversation-Aware Scam Scenario Detection}

Building on scenario-level knowledge learned from large-scale scam reports, we present \sys, a conversation-aware framework for early scam scenario detection in customer-service conversations. As illustrated in Figure~\ref{fig:methodoverview}, \sys consists of a training phase, including Incident Report Pretraining and Transfer Learning for Customer-Service Conversations, and an inference phase. We next describe the threat model and then detail each phase of the framework.

\subsection{Threat Model}
\label{subsec:threatModelProblem}

% \noindent\textbf{Threat Model.}
We consider online scams in which an external attacker crafts deceptive narratives and applies PTs to induce harmful victim actions (e.g., monetary transfer). The attacker communicates via channels such as SMS, email, or social media, often impersonating trusted entities (e.g., agencies, employers, or service providers). We assume a social-engineering setting without technical compromise: the attacker does not gain access to victim devices, banking infrastructure, or communication platforms, but instead relies on scalable template reuse and contextual adaptation to enhance credibility. We further assume no access to the victim's personal data, no direct interaction with the attacker, and exclude data poisoning, infrastructure compromise, automated transaction blocking, and white-box adversaries optimizing against the deployed model.

% \noindent\textbf{Defender.}
The defender is a financial institution or anti-scam service that observes customer-provided narratives or customer-service conversations during fraud investigations. Given partial and evolving information of the incident, the defender aims to infer the underlying scam scenario as early as possible. \autoref{fig:exampleConversation} illustrates an example conversation for the ``\textit{Government}'' scam scenario.

\subsection{Incident Report Pretraining}
\label{subsec:incidentPretrain}

% Although incident reports and customer-service conversations differ in format, they often describe the same underlying scam scenario: conversations reveal the narrative incrementally, while reports provide a retrospective summary. For example, in \autoref{fig:exampleConversation}, the customer discloses key evidence across turns $c_3$–$c_6$, including an urgent utility bill, a shutoff threat, and a gift card payment request, closely matching the incident report in \autoref{fig:exampleUtilityReport}. 

% Incident reports and customer-service conversations often describe the same underlying scam scenario.
% For example, in \autoref{fig:exampleConversation}, the customer discloses key evidence across turns $c_3$–$c_6$, closely matching the incident report in \autoref{fig:exampleUtilityReport}. 
We pretrain a scenario classifier on large-scale incident reports to learn transferable scenario-level semantics for conversations. Because scenario labels are not directly available and require manual annotation of clustering results (as discussed Subsection~\ref{subsec:topicModeling}), we construct the dataset using a two-stage pipeline that combines LLM-based scam content extraction and human verification under active learning.

\noindent\textbf{LLM-Based Scam Content Extraction.}
Incident reports are noisy, often containing emotional complaints, informal language, and irrelevant details, making manual annotation expensive. Prior work shows that LLMs effectively extract scam-relevant content from such data~\cite{ma2025psyscam}. We therefore use an LLM to generate structured draft annotations, prompting it to extract scam evidence, predict the scam scenario, and provide a brief rationale. Although direct LLM-based classification achieves only about $60\%$ accuracy, it reliably surfaces useful evidence for human verification. Thus, we use LLM outputs as annotation assistance rather than final labels. The prompt is provided in Appendix~\autoref{fig:annotationPrompt}.

\noindent\textbf{Multi-Round Active Learning.}
Starting from LLM-assisted annotations, we perform multi-round active learning. In each round, we first uniformly sample an equal number of reports per scenario for human labeling to maintain a balanced dataset. We then expand the labels through similarity-based propagation, using cosine similarity between incident report text embeddings with a threshold of 0.7, which substantially increases coverage while preserving label quality based on spot checks.
Next, we train a scenario classifier  on the expanded set and apply it to the remaining unlabeled data. We select the least confident predictions for human review in the next round, prioritizing informative samples over random selection. After each round, we evaluate performance on the labeled set and adjust the annotation budget based on observed gains.

\noindent\textbf{Data Quality Control.}
The human annotation process involved three PhD-level annotators in computer science. We craft a detailed guideline document defining the taxonomy and labeling protocol, and conduct a calibration stage on a trial set spanning all scenarios to align annotator understanding. Each report is independently labeled by two annotators, and disagreements are resolved by the third annotator, with the final decision determined by majority agreement. The detailed human review protocol is provided in   Appendix~\ref{app:humanreviewprotocol}.

\subsection{Transfer Learning for Customer-Service Conversations}
\label{subsec:synthconv}

We next describe how we conduct the transfer learning that adapts the classifier to the conversational setting.

\noindent\textbf{Formal Definition.}
We represent a customer-service conversation as an ordered sequence of agent--customer
utterance pairs,
\begin{equation}
\mathcal{C} \;=\; \bigl\langle (a_1, c_1),\, (a_2, c_2),\, \ldots,\, (a_T, c_T) \bigr\rangle,
\end{equation}
where $a_t$ and $c_t$ denote the agent and customer utterances at turn
$t$, respectively, and $T$ is the total number of turns.
The \emph{customer narrative} through turn $t$ is the concatenation of
customer-side utterances,
\begin{equation}
\mathcal{N}_t \;=\; c_1 \oplus c_2 \oplus \cdots \oplus c_t.
\end{equation}
We further define two label spaces, i.e., $\mathcal{Y}_{\mathrm{scn}}=\{\textsc{Legitimate}\}\cup\mathcal{S}$ over the scam scenarios $\mathcal{S}$ and $\mathcal{Y}_{\mathrm{tac}}=\{\textsc{Legitimate}\}\cup\mathcal{T}$ over the tactics $\mathcal{T}$.

\noindent\textbf{Phase-Aware Modeling.}
In reality, these customer-service conversations usually unfold in three broad phases:
\textsc{Unrelated} phase contains routine or irrelevant content, such as greetings. \textsc{Narrative} phase contains general background or administrative information, such as identity verification or transaction confirmation. \textsc{Scam\_Evidence} phase contains information useful for identifying the scam scenario, such as the claimed identity of the scammer, the pretext, or the requested payment action.
For example in \autoref{fig:exampleConversation}, $c_1$ is a greeting (\textsc{Unrelated}), $c_2$ provides routine administrative operations (\textsc{Narrative}), and diagnostic evidence appears only from $c_3$ onward (\textsc{Scam\_Evidence}).

To capture this progression, we train a per-turn phase classifier
\begin{equation}
\begin{aligned}
g_\psi:c_t &\mapsto \Delta^{|\mathcal{P}|},\\
\mathcal{P}
&=
\{\textsc{Unrelated},
\textsc{Narrative},
\textsc{Scam\_Evidence}\},
\end{aligned}
\end{equation}
which predicts a phase distribution over $\mathcal{P}$. 
Specifically, we employ a Phase Info Builder to construct two phase-aware signals. First, it converts the predicted phase information into a phase-state feature vector $\phi_t \in \mathbb{R}^{11}$, encoding both the current phase and its progression across turns. Second, to model the temporal proximity to the emergence of scam indicators, it encodes the turn offset $d_t=t-t_e$, where $t_e$ is the onset turn, defined as the first turn labeled as \textsc{Scam_Evidence}.
Details of the phase-state feature vector are provided in Appendix~\autoref{tab:phasefeatures}.

\noindent\textbf{LLM-Based Data Augmentation.}
To better mimic the variability of real customers, we use LLM to mutate each conversation with two complementary augmentation modes.
The first mode performs fine-grained paraphrasing. The LLM rewrites each turn in place, adding spoken style and non-structural lexical variation. The second mode performs coarse-grained conversation-level rewriting. The LLM rewrites the full conversation with varied persona, open style, and narrative structure, while preserving the underlying scenario. An independent LLM judge then checks whether the rewritten conversation preserves the original scenario. Mutations that fail this check are rejected.  
Prompt templates are given in Appendix~\autoref{fig:augmentationPrompts}.
The final training pool is a mixture of source conversations and accepted mutations from both modes.

% This supervision explicitly models the benign-to-scam transition: the model is trained to remain uncommitted (\textsc{Legitimate}) until diagnostic evidence emerges, and to switch to the appropriate tactic and scenario once sufficient signals are observed from the customer narratives. 
% As a result, the model produces a time-indexed sequence of turn-level decisions rather than a single conversation-level label.

\noindent\textbf{Model Architecture.}
Our goal is to train a scenario classifier $f_\theta$ that predicts the scam scenario and tactic at each turn:
\begin{equation}
f_\theta:(\mathcal{N}_t,\phi_t,d_t)\mapsto
\bigl(p_t^{\mathrm{scn}},\,p_t^{\mathrm{tac}}\bigr),
\end{equation}
where $p_t^{\mathrm{scn}}$ and $p_t^{\mathrm{tac}}$ are distributions over scenario and tactic labels given the narrative prefix $\mathcal{N}_t$, phase-state features $\phi_t$, and turn offset $d_t$. 
We initialize $f_\theta$ from the incident-report classifier (\autoref{subsec:incidentPretrain}) and reuse its transformer encoder to encode the narrative prefix:
\begin{equation}
h_t^{\mathrm{inc}} = e_{\mathrm{inc}}(\mathcal{N}_t).
\end{equation}
We make each prefix encoding position-aware with a learnable turn-position embedding $e_{\mathrm{pos}}(t)$, giving the base conversation representation
\begin{equation}
\bar{h}_t = h^{\mathrm{inc}}_t + e_{\mathrm{pos}}(t),
\end{equation}
which is also used by the no-phase baseline. \sys then injects the two phase signals into the same hidden space through independent parameters:
\begin{equation}
h_t = \bar{h}_t + W_{\phi}\phi_t + e_{\mathrm{off}}(d_t),
\end{equation}
where $W_{\phi}$ is a learnable projection of the phase-state vector $\phi_t$ and $e_{\mathrm{off}}(\cdot)$ is a learnable embedding of the evidence offset $d_t$. The final hidden representation $h_t$ is fed into two parallel heads that predict the scam scenario and tactic:
\begin{equation}
p^{\mathrm{scn}}_t = \mathrm{softmax}(W_{\mathrm{scn}}h_t), \qquad p^{\mathrm{tac}}_t = \mathrm{softmax}(W_{\mathrm{tac}}h_t).
\end{equation}
This design transfers scenario-level semantics learned from incident reports while adapting to the turn-by-turn structure of customer-service conversations.

\noindent\textbf{Training.}
    As discussed above, customer-service conversations progress through three phases. During the \textsc{Unrelated} and \textsc{Narrative} phases, \sys outputs \textsc{Legitimate}; once the conversation enters the \textsc{Scam_Evidence} phase, \sys predicts the corresponding scam scenario and tactic labels.
    To model this setting, we design a phase-aware turn loss (\textsc{PATL}) for training the scenario and tactic classifier. Let $p_t$ be the predicted  distribution at turn $t$, $y_t \in \mathcal{Y}=\{\textsc{Legitimate}\}\cup\mathcal{Y}_{\mathrm{scn}}$ be the classification label, and $\mathcal{V}$ be the set of supervised turns. The loss is defined as 
\begin{equation} \mathcal{L}_{\textsc{patl}}(p, y)= \frac{\sum_{t\in\mathcal{V}} w_t\,\alpha_{y_t}\,\mathrm{CE}(p_t,y_t)} {\sum_{t\in\mathcal{V}} w_t}, 
\end{equation} 
where $w_t$ is a turn weight and $\alpha_{y_t}$ an inverse-frequency class weight (balancing power $0.5$) that supports long-tail classes. We up-weight the decisive turns: $w_t=2.5$ at the first \textsc{Scam\_Evidence} onset turn, $w_t=1.5$ for other \textsc{Scam\_Evidence} and \textsc{Narrative} turns, and $w_t=1.0$ for \textsc{Unrelated} turns. The scenario and tactic heads are optimized jointly with the combined objective:  
\begin{equation}
\mathcal{L}=\mathcal{L}_{\textsc{patl}}(p^{\mathrm{scn}}, y^{\mathrm{scn}})+\mathcal{L}_{\textsc{patl}}(p^{\mathrm{tac}}, y^{\mathrm{tac}}), 
\end{equation}
encouraging the model to focus on diagnostic evidence while reducing bias toward frequent classes.

% $\mathcal{P}=g_\psi(c_t)$. If $\mathcal{P}=\textsc{Scam_Evidence}$.

\noindent\textbf{Inference.}
During live customer-service conversations, \sys processes turns incrementally. At current turn $t$, \sys first applies the phase classifier $g_\psi$ to the current customer utterance $c_t$ to determine the phase, $\mathcal{P}=g_\psi(c_t)$. If $\mathcal{P}=\textsc{Scam_Evidence}$, the current turn is marked as the \textsc{Scam_Evidence} onset turn, $\hat{t}$. For every subsequent turn $t>\hat{t}$, the utterance $c_t$ is passed to the scenario classifier.
\sys then employs the Phase Info Builder to construct the phase-state feature vector $\phi_t$ from $c_t$ and the narrative prefix $\mathcal{N}_t$ (i.e., $c_1 \oplus c_2 \oplus \cdots \oplus c_t$), and predicts the scenario and tactic distributions, $p^{\mathrm{scn}}_t$ and $p^{\mathrm{tac}}_t$, respectively.
\sys then emits the scenario and tactic prediction: 
\begin{equation}
    \hat{y}^{\mathrm{scn}}_t=\arg\max_y p^{\mathrm{scn}}_t(y), \qquad \hat{y}^{\mathrm{tac}}_t=\arg\max_y p^{\mathrm{tac}}_t(y).
\end{equation}
The system alerts the customer-service agent with the predicted scenario and tactic to support timely intervention, such as at $a_6$ in \autoref{fig:exampleConversation} and $c_3$ in \autoref{fig:methodoverview}. 
%Otherwise, it waits for additional turns and continuously updates its prediction as more customer evidence becomes available.

\section{Evaluation on Scam Scenario Detection}

% \subsection{Evaluation Setup}
% \noindent\textbf{Dataset.}
% Our dataset is obtained from a security company.
% Due to privacy issue, it impratical to directly share user data.   synthesized dataset.
% Real customer service conversations are private, low-volume, and expensive to
% annotate, making it difficult to train \(f_\theta\) directly using
% large-scale in-domain data.
% \noindent\textbf{Data Availability.}
% A limitation of our evaluation is that it relies on synthetic conversations provided by our industry partner. Real-world banking customer conversations are highly sensitive and cannot be directly shared with us due to privacy concerns. However, such data would be available within financial institutions in practical deployments. In future work, we plan to collaborate with the company to deploy our model internally and evaluate it on real customer conversations under strict privacy-preserving protocols.\xiao{move this to eval setup and tone it down}

\subsection{Evaluation Setup}

\noindent\textbf{Dataset.}
We evaluate our approach on a customer-service conversation dataset provided by Charm Security, a leading security company focused on fraud and cybercrime. Because real customer conversations are subject to strict privacy and data-sharing restrictions, the partner generated synthetic conversations based on real fraud-investigation calls. This choice reflects institutional privacy constraints rather than a deployment limitation, as the detector can be applied directly to institution-internal conversations without exposing customer data externally.  The dataset contains $1{,}115$ customer-service conversations, with 948 for training, and 167 for testing. Every turn carries both a tactic label and a scenario label.
The per-scenario conversation counts and turn statistics are reported in \autoref{tab:convdataset}.

Our data augmentation further expands the original 948 training set into $2{,}277$ conversations. % ($29{,}630$ turns).
Specifically, the fine-grained mode produces $7{,}061$ accepted variants and $523$ rejected variants, yielding a $93.1\%$ acceptance rate; the coarse-grained mode produces $4{,}590$ accepted variants and $2{,}994$ rejected variants, yielding a $60.5\%$ acceptance rate. %Unless otherwise stated, our main training configuration augments each source training conversation with two accepted variants: one fine-grained mutation and one coarse-grained mutation. 

\noindent\textbf{Implementation.}
We implement \sys with three state-of-the-art transformer encoders, RoBERTa~\cite{liu2019roberta}, ELECTRA~\cite{clark2020electra}, and DeBERTa~\cite{hedeberta}, and report results for each backbone. Owing to the limited size of the conversation training data and the cost of full fine-tuning, we adopt the base-size variant of each encoder (hidden size $768$). Each fold follows the same two-stage recipe: incident-report pretraining (\autoref{subsec:incidentPretrain}) initializes the backbone, and conversation fine-tuning then trains the turn-position embedding, phase-conditioned projection, and two task heads. Fine-tuning uses AdamW with weight decay $0.01$, encoder learning rate $1\times10^{-5}$, head/injection learning rate $8\times10^{-5}$, batch size $1$ with $4$ gradient-accumulation steps (effective batch $4$), and $10$ epochs; the checkpoint is selected by the validation multi-task score. Inputs are truncated to $\texttt{max\_turns}=64$ turns and a $512$-token prefix budget. 
The phase classifier is trained with weighted cross entropy and AdamW, using a learning rate $2\times10^{-5}$, maximum length $256$, batch size $16$, and $5$ epochs. 

\noindent\textbf{Evaluation Metrics.} 
We report the accuracy and macro-F1 to measure the classification performance under 5-fold cross-validation. 

% For incident-report pretraining,  we report scenario and tactic top-1 accuracy and macro-F1, while the active-learning analysis focuses on top-1 accuracy as a function of annotation round.
% For transfer learning to customer-service conversations, pre-evidence turns are intentionally labeled \textsc{Legitimate} and primarily serve to evaluate detection timing. Therefore, our primary evaluation universe consists of post-evidence turns ($t \ge t_e$), where $t_e$ denotes the annotated evidence onset. 
% % This setting isolates scenario-discrimination performance once diagnostic evidence becomes available. 
% % Unless otherwise stated, turn-level metrics are computed by pooling post-evidence turns from the fixed 167-conversation test set within each fold and averaging results across five folds.
% We report scenario and tactic macro-F1 as the primary metrics, averaging over all classes present in the post-evidence ground truth. The \textsc{Legitimate} class is excluded from the averaging procedure, although incorrect predictions of \textsc{Legitimate} are still counted as errors. We additionally report scenario top-1 and top-3 accuracy.
% To quantify earliness, we measure scenario top-1 accuracy as a function of the relative turn offset $d=t-t_e$, where $d=0$ corresponds to the first evidence-bearing turn. For each fold, turns sharing the same offset are pooled before computing accuracy, and the resulting curves are averaged across folds.
% All the results are reported under $5$-fold cross-validation.

\noindent\textbf{Research Questions.}
We aim to answer the following research questions:
\begin{itemize}[noitemsep, topsep=1pt, partopsep=1pt, listparindent=\parindent, leftmargin=*]
\item \textbf{RQ1:} How effective is active learning for incident report pretraining?
\item \textbf{RQ2:} How effective is \sys in detecting scam scenarios in customer-service conversations?  
\item \textbf{RQ3:} How does each component contribute to \sys's overall performance?
\end{itemize}

\subsection{RQ1: Incident Reports Pretraining}

\begin{table}[t]
\centering
\small
\setlength{\tabcolsep}{4pt}
\renewcommand{\arraystretch}{1.1}
\caption{Dataset size across active learning rounds.}
\resizebox{\columnwidth}{!}{
\begin{tabular}{lccc}
\toprule
 & \textbf{Round 1} & \textbf{Round 2} & \textbf{Round 3} \\
\midrule
Human-annotated reports     &  1,523 &  2,202 &  2,413 \\
Similarity-expanded reports  & 28,710 & 36,527 & 37,569 \\
\bottomrule
\end{tabular}
}
\label{tab:activeLearningResults}
\end{table}

\begin{figure}[t]
\centering
% --- shared axis style for both subplots ---
\pgfplotsset{
    accConvergence/.style={
        width=\linewidth,
        height=0.85\linewidth,
        xlabel={Active Learning Rounds},
        ylabel={Accuracy (\%)},
        xmin=0.7, xmax=3.3,
        xtick={1,2,3},
        xticklabels={R1, R2, R3},
        legend style={
            font=\tiny,
            draw=black!60,
            fill=white,
            fill opacity=0.92,
            text opacity=1,
            row sep=0pt,
            inner sep=2pt,
            at={(0.98,0.04)}, anchor=south east,
        },
        legend cell align={left},
        grid=major,
        grid style={dashed, black!15},
        tick align=outside,
        tick pos=left,
        axis line style={black!70},
        label style={font=\footnotesize},
        tick label style={font=\scriptsize},
        every axis plot/.append style={line width=0.9pt},
    }
}

% ---------- Category ----------
\begin{subfigure}[t]{0.48\linewidth}
\centering
\begin{tikzpicture}
\begin{axis}[
    accConvergence,
    ymin=82, ymax=92,
    ytick={83,85,87,89,91},
]
\addplot[color=blue!70!black, dashed, no marks, domain=0.7:3.3] {83.98};
\addlegendentry{LLM}
\node[font=\tiny, color=blue!70!black, anchor=south]
    at (axis cs:1.2, 83.98) {83.98};
\addplot[
    color=blue!70!black, mark=*, mark size=2pt,
    mark options={fill=blue!70!black},
    nodes near coords,
    every node near coord/.append style={
        font=\tiny,
        color=blue!70!black,
        anchor=south,
        yshift=1pt,
        /pgf/number format/.cd, fixed, fixed zerofill, precision=2,
    },
]
coordinates {(1,87.32) (2,91.28) (3,91.67)};
\addlegendentry{DeBERTa}
\end{axis}
\end{tikzpicture}
\caption{Tactic}
\label{fig:scenarioCategory}
\end{subfigure}
\hfill
% ---------- Subcategory ----------
\begin{subfigure}[t]{0.48\linewidth}
\centering
\begin{tikzpicture}
\begin{axis}[
    accConvergence,
    ymin=64, ymax=86,
    ytick={66,70,74,78,82},
]
\addplot[color=red!70!black, dashed, no marks, domain=0.7:3.3] {67.50};
\addlegendentry{LLM}
\node[font=\tiny, color=red!70!black, anchor=south]
    at (axis cs:1.2, 67.50) {67.50};
    
\addplot[
    color=red!70!black, mark=square*, mark size=1.8pt,
    mark options={fill=red!70!black},
    nodes near coords,
    every node near coord/.append style={
        font=\tiny,
        color=red!70!black,
        anchor=south,
        yshift=1pt,
        /pgf/number format/.cd, fixed, fixed zerofill, precision=2,
    },
]
    coordinates {(1,79.81) (2,83.62) (3,84.38)};
\addlegendentry{DeBERTa}
\end{axis}
\end{tikzpicture}
\caption{Scenario}
\label{fig:scenarioSubcategory}
\end{subfigure}

\caption{Pretraining performance across active learning rounds.
Dashed horizontal lines denote direct LLM classification.}
\label{fig:activeLearningResults}
\end{figure}
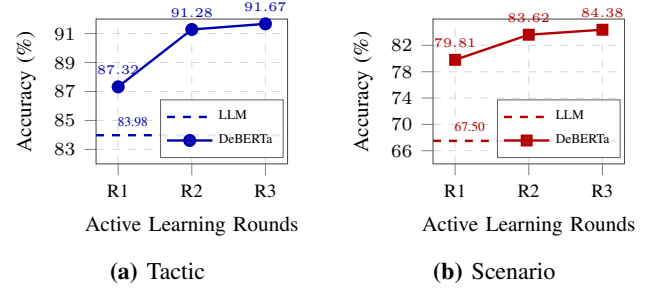

\begin{table}[t]
\centering
\small
\setlength{\tabcolsep}{2pt}
\caption{Incident reports pretraining performance.}
\resizebox{\columnwidth}{!}{
\begin{tabular}{lcccc}
\toprule
\textbf{Model} & \textbf{Tactic Acc.} & \textbf{Scenario Acc.} & \textbf{Tactic F1} & \textbf{Scenario F1} \\
\midrule
RoBERTa  & $90.93_{\pm 2.13}$ & $81.69_{\pm 4.59}$ & $90.99_{\pm 2.33}$ & $79.21_{\pm 6.74}$ \\
ELECTRA & $\mathbf{91.80}_{\pm \mathbf{1.72}}$ & $81.82_{\pm 3.10}$ & $92.01_{\pm 1.80}$ & $81.18_{\pm 4.13}$ \\
DeBERTa  & $91.67_{\pm 1.34}$ & $\mathbf{84.38}_{\pm \mathbf{1.47}}$ & $\mathbf{92.05}_{\pm \mathbf{1.52}}$ & $\mathbf{85.23}_{\pm \mathbf{1.43}}$ \\
\bottomrule
\end{tabular}
}
%TODO: Do we need classical ML models such as Logistic regression, Random Forest
\label{fig:scenarioBBBResults}
\end{table}

\sys employs multi-round active learning to improve pretraining performance.
\autoref{tab:activeLearningResults} summarizes dataset growth across three active-learning rounds, and \autoref{fig:activeLearningResults} reports the corresponding pretraining performance. The number of human-annotated reports increases from 1,523 to 2,413, while similarity-based expansion enlarges the training set from 28,710 to 37,569 reports. As the training corpus grows, tactic accuracy improves from $87.32\%$ to $89.33\%$ and scenario accuracy from $79.81\%$ to $82.74\%$, substantially outperforming direct LLM classification ($83.98\%$ and $67.50\%$, respectively). 
Performance gains largely saturate after Round 2, with tactic accuracy increasing by only $0.05$ percentage points in Round 3 ($89.28\% \rightarrow 89.33\%$). We therefore stop the active-learning process after the third round.

 \autoref{fig:scenarioBBBResults} shows the performance of models trained on the final 37,569 incident reports. All models achieve strong tactic-classification performance, exceeding $90\%$ accuracy and F1, indicating that high-level scam tactics are readily distinguishable from incident reports. In contrast, scenario classification is more challenging because of the greater diversity among scenarios; nevertheless, the model still achieves strong performance.   
 DeBERTa achieves the best results, with $84.38\%$ scenario accuracy, $85.23\%$ scenario F1, and $92.05\%$ tactic F1. These results validate the effectiveness of our active learning pipeline and establish a strong foundation for transferring scenario-level knowledge to conversational scam detection. Detailed results  across all scenarios and tactics are reported in Appendix~\autoref{tab:scenarioSubcategoryResults}.

\subsection{RQ2: Customer-Service Conversations}
\label{subsec:rq2}

\begin{table}[t]
\centering
\scriptsize
\setlength{\tabcolsep}{3pt}
\caption{Tactic classification results on customer-service conversations. }
\label{tab:rq2tactic}
\begin{tabular}{llcc}
\toprule
\textbf{Method} & \textbf{Variants} & \textbf{Tactic F1} & \textbf{Tactic Acc} \\
\midrule
\multirow{4}{*}{LLM Zero-shot}
& Llama-3.1-8B & $38.33$ & $46.48$ \\
& GPT-OSS-20B & $51.84$ & $58.03$ \\
& Qwen3-32B & $53.51$ & $59.16$ \\
& GPT-5.4 & $64.87$ & $67.12$ \\
\cmidrule(l){1-4}
\multirow{4}{*}{LLM Few-shot}
& Llama-3.1-8B & $62.11_{\pm5.49}$ & $63.07_{\pm4.53}$ \\
& GPT-OSS-20B & $64.51_{\pm0.79}$ & $66.97_{\pm0.71}$ \\
& Qwen3-32B & $66.97_{\pm2.36}$ & $68.70_{\pm1.92}$ \\
& GPT-5.4 & $70.94_{\pm2.64}$ & $72.13_{\pm2.60}$ \\
\midrule
\multirow{3}{*}{Transformer}
& RoBERTa & $81.32_{\pm1.62}$ & $82.09_{\pm1.45}$ \\
& ELECTRA & $80.67_{\pm2.45}$ & $81.21_{\pm2.23}$ \\
& DeBERTa & $80.92_{\pm1.40}$ & $81.34_{\pm1.24}$ \\
\midrule
\multirow{3}{*}{\sys}
& RoBERTa & $83.35_{\pm1.20}$ & $83.80_{\pm1.11}$ \\
& ELECTRA & $\mathbf{84.14}_{\pm1.57}$ & $\mathbf{84.41}_{\pm1.60}$ \\
& DeBERTa & $83.80_{\pm0.54}$ & $83.97_{\pm0.66}$ \\
\bottomrule
\end{tabular}
\end{table}

\begin{table}[t]
\centering
\scriptsize
\setlength{\tabcolsep}{2pt}
\caption{Scenario classification results on customer-service conversations.  Acc@3 counts a turn as correct if the ground-truth scenario appears among the model's top-3 predicted scenarios.}
\label{tab:rq2scenario}
\begin{tabular}{llccc}
\toprule
\textbf{Method} & \textbf{Variants} & \textbf{Scn F1} & \textbf{Scn Acc} & \textbf{Acc@3} \\
\midrule
\multirow{4}{*}{LLM Zero-shot}
& Llama-3.1-8B & $15.56$ & $32.15$ & $47.52$ \\
& GPT-OSS-20B & $29.44$ & $44.11$ & $57.48$ \\
& Qwen3-32B & $33.15$ & $45.88$ & $68.12$ \\
& GPT-5.4 & $42.52$ & $52.61$ & $78.35$ \\
\cmidrule(l){1-5}
\multirow{4}{*}{LLM Few-shot}
& Llama-3.1-8B & $30.74_{\pm3.80}$ & $41.52_{\pm2.75}$ & $65.23_{\pm4.19}$ \\
& GPT-OSS-20B & $34.59_{\pm1.18}$ & $47.29_{\pm1.25}$ & $68.62_{\pm1.90}$ \\
& Qwen3-32B & $42.14_{\pm3.02}$ & $50.84_{\pm2.01}$ & $73.07_{\pm3.38}$ \\
& GPT-5.4 & $45.77_{\pm1.89}$ & $54.23_{\pm1.61}$ & $81.90_{\pm1.01}$ \\
\midrule
\multirow{3}{*}{Transformer}
& RoBERTa & $66.15_{\pm2.35}$ & $71.39_{\pm2.26}$ & $88.88_{\pm1.87}$ \\
& ELECTRA & $64.89_{\pm2.70}$ & $70.13_{\pm2.36}$ & $87.71_{\pm1.57}$ \\
& DeBERTa & $65.39_{\pm1.38}$ & $71.21_{\pm1.85}$ & $\mathbf{91.64}_{\pm1.42}$ \\
\midrule
\multirow{3}{*}{\sys}
& RoBERTa & $70.33_{\pm1.68}$ & $73.72_{\pm1.37}$ & $89.81_{\pm1.07}$ \\
& ELECTRA & $\mathbf{71.39}_{\pm1.60}$ & $74.96_{\pm2.04}$ & $88.61_{\pm1.16}$ \\
& DeBERTa & $69.83_{\pm2.01}$ & $\mathbf{75.17}_{\pm1.25}$ & $91.04_{\pm0.94}$ \\
\bottomrule
\end{tabular}
\end{table}

\noindent\textbf{Baselines.}
We compare \sys against three groups of baselines. First, \textit{LLM Zero-shot} prompts an LLM with the scenario taxonomy and the customer-service conversation, and asks it to predict the underlying tactic and scenario without task-specific examples. Second, \textit{LLM Few-shot} uses the same prompt setting but includes five labeled examples with classification rationales.  
% The prompts are shown in Appendix~\autoref{fig:llmBaselinePrompt}. 
Third, \textit{Transformer} fine-tunes off-the-shelf transformer encoders and a phase classifier on the conversation dataset without incorporating \sys's data augmentation. To ensure a fair comparison, we adopt the same inference procedure as \sys (\autoref{subsec:synthconv}): the transformer encoder is invoked only after the phase classifier identifies an onset turn; otherwise, all preceding turns are treated as legitimate.

\noindent\textbf{Overall Effectiveness.}
\autoref{tab:rq2tactic} and \autoref{tab:rq2scenario} show the overall performance of \sys in tactic and scenario classification.  
\sys achieves the best macro-F1 and accuracy on both tasks. For tactic classification, \sys achieves $84.14\%$ macro-F1 and $84.41\%$ accuracy, improving over the best Transformer baseline by $2.82$ and $2.32$ points, and over the best few-shot LLM by $13.20$ and $12.28$ points. 
 For scenario classification, \sys obtains up to $71.39\%$ macro-F1 and $75.17\%$ accuracy, outperforming the best Transformer baseline by $5.24$ and $3.78$ percentage points, respectively. The improvement is more substantial over LLM baselines: compared with the best few-shot LLM, \sys improves scenario macro-F1 by $25.62$ points and accuracy by $20.94$ points. Beyond higher means, \sys is also more stable across folds: its standard deviations are lower than the Transformer baseline's on nearly all metrics.
 These results show that directly prompting LLMs remains insufficient for reliable conversation-level scam understanding, while \sys effectively transfers scenario knowledge to customer-service conversations through augmentation and phase-aware modeling. 
For Acc@3, \sys achieves performance comparable to the Transformer  baselines, while maintaining a slight overall advantage. Although DeBERTa attains a marginally higher mean Acc@3, it exhibits greater variance across folds. This is likely because each tactic is associated with roughly three scenarios on average; once the correct tactic is identified, the correct scenario is often included among the top three predictions, making Acc@3 less sensitive to fine-grained scenario ranking differences.

 % It is also worth noting that the Transformer baseline with DeBERTa achieves a slightly higher Acc@3 than \sys but with higher standard deviation across folds. The gap is within one standard deviation across folds, and .

The roughly ten-point gap between \sys's tactic and scenario classification performance is itself informative. Based on our manual inspection of the classification errors, nearly half still preserve the correct tactic, as the model often confuses closely related scenarios within the same tactic (e.g., \emph{fake job offer} vs.\ \emph{unpaid labor} under \emph{employment scam}). These near-misses are well covered by the top-3 predictions, which contain the correct scenario for nearly $90\%$ of these sibling confusions and reach up to $91.04\%$ Acc@3. Therefore, surfacing the top-3 candidate scenarios provides useful triage support and better reflects the model's ability to handle the taxonomy's ambiguity.
 
% reaches only $41.7$ scenario macro-F1, compared with $72.1$ for \sys, leaving a roughly $30$-point gap that also persists for tactic macro-F1. Larger models and in-context demonstrations help, but they do not close this gap. The shortfall is not a formatting artifact, since fewer than $1\%$ of outputs are unparseable or out-of-taxonomy. Instead, it reflects genuine task difficulty: errors are dominated by confident misclassification and by under-detection at post-evidence turns. This echoes our annotation-stage finding that direct LLM scenario labeling is weak, with about $60\%$ accuracy, and indicates that conversation-aware fine-tuning is what drives scenario detection.

% \noindent\textbf{Comparison with LLMs.}
%We prompt each LLM under the same causal, per-prefix protocol as \sys , reporting zero-shot and few-shot results. The few-shot setting uses five training-pool demonstration sets shared across all models. Across all three model sizes, prompting remains far below the fine-tuned detector, even with few-shot prompting at $32$B parameters. 

\noindent\textbf{Late vs.\ Early Correctness.}
Our goal is to detect scam scenarios as early as possible, but early prediction naturally trades off with accuracy: as the conversation progresses, customers reveal more scam evidence, allowing the model to make more confident and accurate predictions.
To quantify this tradeoff, we evaluate \sys at different turns and measure how scenario-classification accuracy changes over time. Specifically, we define the turn offset as $d=t-t_e$, where $d=0$ denotes the first turn in the \textsc{Scam_Evidence} phase. \autoref{fig:peroffset} reports scenario classification accuracy with varying offset.

Overall, accuracy increases steadily as the offset grows. At the onset turn ($d=0$), \sys is less accurate ($\sim$56--58\%), which is expected because limited evidence is available at this point to enable accurate classification. Nevertheless, a single additional turn is decisive: at $d=1$, scenario accuracy improves by $11$--$16$ percentage points and already reaches the overall accuracy. Subsequent gains are smaller but consistent, reaching around $78$--$79\%$ by $d=6$. These results show that \sys typically identifies the correct scenario within one to two turns after scam evidence first appears, supporting early intervention while still benefiting from additional evidence as the conversation unfolds.

% \sys holds a steady $+3$ to $+4$ point scenario-accuracy lead over the augmentation-only baseline across all post-onset offsets ($d\ge1$) on ELECTRA and DeBERTa, and matches or exceeds it on RoBERTa. 
% At the onset turn ($d=0$) the deployable gain over augmentation-only is more modest ($+3.95$ on ELECTRA, $+1.08$ on DeBERTa, $+0.60$ on RoBERTa), because the system must \emph{infer} the onset rather than observe; a much larger onset spike is attainable only when the offset is anchored on the gold onset (e.g., $+10.30$ on ELECTRA), which we treat as an oracle and do not report as a deployable result. Augmentation alone already helps mid- and late-conversation turns but slightly \emph{hurts} the onset turn on RoBERTa ($-2.75$) and DeBERTa ($-4.91$); phase injection recovers the onset and then sustains the lead. Together, augmentation broadens context coverage while phase injection provides a consistent post-onset accuracy lift.

% \begin{figure}[t]
% \centering
% \includegraphics[width=\columnwidth]{figs/rq1PerOffsetAccuracy.pdf}
% \caption{Per-turn scenario accuracy vs.\ offset from the evidence onset
% ($\text{offset}=t-t_e$).}
% \label{fig:peroffset}
% \end{figure}

\pgfplotsset{compat=1.16}
\usepgfplotslibrary{groupplots}
\tikzset{acclbl/.style={inner sep=0.5pt}}
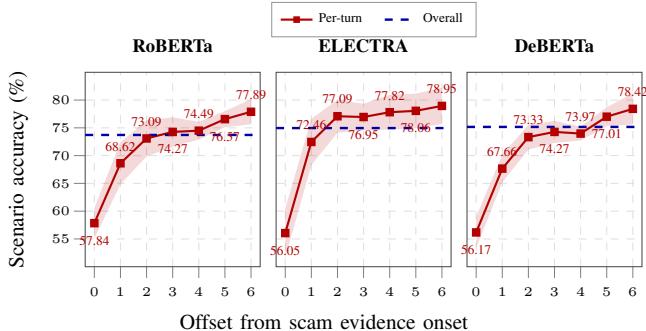
\begin{figure}[t]
\centering
\pgfplotsset{
    perOffset/.style={
        scale only axis=true,
        width=0.27\columnwidth,
        height=0.30\columnwidth,
        xmin=-0.35,
        xmax=6.35,
        xtick={0,1,2,3,4,5,6},
        ymin=50,
        ymax=85,
        ytick={55,60,65,70,75,80},
        grid=major,
        grid style={dashed, black!12},
        tick align=outside,
        tick pos=left,
        axis line style={black!70},
        label style={font=\footnotesize},
        tick label style={font=\tiny},
        ylabel style={font=\footnotesize, yshift=2pt},
        yticklabel style={xshift=2pt},
        title style={font=\scriptsize\bfseries, yshift=-2pt},
        clip=false,
    },
    meanline/.style={
        color=red!70!black,
        mark=square*,
        mark size=1.2pt,
        mark options={fill=red!70!black},
        line width=0.8pt
    },
    stdband/.style={
        fill=red!70!black,
        opacity=0.15,
        draw=none
    },
    overallline/.style={
        color=blue!65!black,
        dashed,
        line width=0.9pt
    },
}

\begin{tikzpicture}
\begin{groupplot}[
    perOffset,
    group style={
        group size=3 by 1,
        horizontal sep=6pt,
        yticklabels at=edge left
    },
]

% ---- RoBERTa ----
\nextgroupplot[
    title={RoBERTa},
    ylabel={Scenario accuracy (\%)},
    legend style={
        at={(1.65,1.18)},
        anchor=south,
        legend columns=2,
        font=\tiny,
        draw=black!40,
        /tikz/every even column/.append style={column sep=6pt}
    },
    legend image post style={scale=0.7}
]
\addplot[stdband, forget plot] coordinates {
(0,60.60)(1,72.40)(2,76.34)(3,76.91)(4,76.09)(5,78.02)(6,80.16)
(6,75.62)(5,75.12)(4,72.89)(3,71.63)(2,69.84)(1,64.84)(0,55.08)
} \closedcycle;
\addplot[meanline] coordinates {
(0,57.84)(1,68.62)(2,73.09)(3,74.27)(4,74.49)(5,76.57)(6,77.89)
};
\addlegendentry{Per-turn}
\addplot[overallline] coordinates {(-0.35,73.72)(6.35,73.72)};
\addlegendentry{Overall}

\node[acclbl, font=\tiny, color=red!70!black, anchor=north, yshift=-4pt] at (axis cs:0,57.84) {57.84};
\node[acclbl, font=\tiny, color=red!70!black, anchor=south, yshift=4pt] at (axis cs:1,68.62) {68.62};
\node[acclbl, font=\tiny, color=red!70!black, anchor=south, yshift=4pt] at (axis cs:2,73.09) {73.09};
\node[acclbl, font=\tiny, color=red!70!black, anchor=north, yshift=-4pt] at (axis cs:3,74.27) {74.27};
\node[acclbl, font=\tiny, color=red!70!black, anchor=south, yshift=4pt] at (axis cs:4,74.49) {74.49};
\node[acclbl, font=\tiny, color=red!70!black, anchor=north, yshift=-4pt] at (axis cs:5,76.57) {76.57};
\node[acclbl, font=\tiny, color=red!70!black, anchor=south, yshift=4pt] at (axis cs:6,77.89) {77.89};

% ---- ELECTRA ----
\nextgroupplot[
    title={ELECTRA},
    ytick style={draw=none}
]
\addplot[stdband] coordinates {
(0,60.50)(1,76.63)(2,80.07)(3,79.33)(4,80.95)(5,81.16)(6,82.23)
(6,75.67)(5,74.96)(4,74.69)(3,74.57)(2,74.11)(1,68.29)(0,51.60)
} \closedcycle;
\addplot[meanline] coordinates {
(0,56.05)(1,72.46)(2,77.09)(3,76.95)(4,77.82)(5,78.06)(6,78.95)
};
\addplot[overallline] coordinates {(-0.35,74.96)(6.35,74.96)};

\node[acclbl, font=\tiny, color=red!70!black, anchor=north, yshift=-4pt] at (axis cs:0,56.05) {56.05};
\node[acclbl, font=\tiny, color=red!70!black, anchor=south, yshift=4pt] at (axis cs:1,72.46) {72.46};
\node[acclbl, font=\tiny, color=red!70!black, anchor=south, yshift=4pt] at (axis cs:2,77.09) {77.09};
\node[acclbl, font=\tiny, color=red!70!black, anchor=north, yshift=-4pt] at (axis cs:3,76.95) {76.95};
\node[acclbl, font=\tiny, color=red!70!black, anchor=south, yshift=4pt] at (axis cs:4,77.82) {77.82};
\node[acclbl, font=\tiny, color=red!70!black, anchor=north, yshift=-4pt] at (axis cs:5,78.06) {78.06};
\node[acclbl, font=\tiny, color=red!70!black, anchor=south, yshift=4pt] at (axis cs:6,78.95) {78.95};

% ---- DeBERTa ----
\nextgroupplot[
    title={DeBERTa},
    ytick style={draw=none}
]
\addplot[stdband] coordinates {
(0,59.02)(1,70.13)(2,75.56)(3,76.26)(4,75.22)(5,78.71)(6,81.24)
(6,75.60)(5,75.31)(4,72.72)(3,72.28)(2,71.10)(1,65.19)(0,53.32)
} \closedcycle;
\addplot[meanline] coordinates {
(0,56.17)(1,67.66)(2,73.33)(3,74.27)(4,73.97)(5,77.01)(6,78.42)
};
\addplot[overallline] coordinates {(-0.35,75.17)(6.35,75.17)};

\node[acclbl, font=\tiny, color=red!70!black, anchor=north, yshift=-4pt] at (axis cs:0,56.17) {56.17};
\node[acclbl, font=\tiny, color=red!70!black, anchor=south, yshift=4pt] at (axis cs:1,67.66) {67.66};
\node[acclbl, font=\tiny, color=red!70!black, anchor=south, yshift=4pt] at (axis cs:2,73.33) {73.33};
\node[acclbl, font=\tiny, color=red!70!black, anchor=north, yshift=-4pt] at (axis cs:3,74.27) {74.27};
\node[acclbl, font=\tiny, color=red!70!black, anchor=south, yshift=4pt] at (axis cs:4,73.97) {73.97};
\node[acclbl, font=\tiny, color=red!70!black, anchor=north, yshift=-4pt] at (axis cs:5,77.01) {77.01};
\node[acclbl, font=\tiny, color=red!70!black, anchor=south, yshift=4pt] at (axis cs:6,78.42) {78.42};

\end{groupplot}
\end{tikzpicture}\\[-2pt]

{\footnotesize Offset from scam evidence onset}

\caption{Per-turn scenario classification accuracy of \sys.}
\label{fig:peroffset}
\end{figure}
% Dataset: 7 rows, 4 cols (acc,f1,p,r)
% - LLM: zero shot, few shot, different backbones
% -  Deberta, roberta, electra

% overall table
% late detection

\subsection{RQ3: Ablations}
\label{subsec:ablation}

In this part, we ablate two techniques used for transferring the incident-report knowledge to customer-service conversations: LLM-based data augmentation and phase-aware modeling. \autoref{tab:ablchain} shows that both techniques consistently improve scenario-classification F1 across all three backbone models. Starting from the baseline without augmentation or phase modeling, adding LLM-based augmentation improves F1 by $2.84\%$, $4.18\%$, and $2.17\%$ for ELECTRA, RoBERTa, and DeBERTa, respectively. This suggests that augmented conversations help bridge the distribution gap between incident reports and customer-service dialogues by exposing the model to more diverse conversational expressions of the same scam scenarios. Phase-aware modeling further improves performance on top of augmentation, with additional gains of $3.68\%$, $0.34\%$, and $2.44\%$ for ELECTRA, RoBERTa, and DeBERTa, respectively. The improvement indicates that explicitly modeling conversation phases helps the classifier focus on scenario-relevant turns, rather than treating all turns equally.

\noindent\textbf{Data Augmentation Ratio.}
\autoref{tab:augratio} reports the performance of \sys under different augmentation ratios without phase conditioning.
Overall, data augmentation improves scenario macro-F1 for all backbones, with gains over the no-augmentation baseline ranging from $3.26\%$ to $4.18\%$ at their best settings.
The optimal ratio varies across backbones: ELECTRA achieves its best F1 at $8\times$ ($68.13\%$), RoBERTa peaks at $2\times$ ($70.00\%$), and DeBERTa performs best at $4\times$ ($69.08\%$).
However, larger augmentation ratios do not always yield further gains, as performance drops for RoBERTa at $4\times$ and for DeBERTa at $8\times$.
This suggests that moderate augmentation is generally effective, while excessive augmentation may introduce noise or redundant conversational patterns.

\noindent\textbf{Phase Input at Inference Time.}
\autoref{tab:phaseinput} replace \sys's phase classifier with an oracle setting that uses the ground-truth phase label at inference time. 
We can see that oracle phase labels consistently improve performance across all backbones, with moderate gains: scenario F1 increases by $2.69\%$, $2.97\%$, and $2.27\%$ for ELECTRA, RoBERTa, and DeBERTa, respectively.
Similarly, scenario accuracy improves by $4.29\%$, $4.55\%$, and $3.67\%$.
These gaps indicate that the learned phase classifier already supplies useful phase information, while also showing that more accurate phase prediction could further improve scenario detection.

\begin{table}[t]
\centering
\small
\setlength{\tabcolsep}{1.5pt}
\caption{Ablation study of context-aware scenario detection. Results report scenario-classification F1; values in parentheses indicate the improvement over the previous row.}

\label{tab:ablchain}
\resizebox{\columnwidth}{!}{
\begin{tabular}{lccc}
\toprule
\textbf{Added factor} & \textbf{ELECTRA} & \textbf{RoBERTa} & \textbf{DeBERTa} \\
\midrule
Baseline (no aug., no phase) & $64.87$ & $65.82$ & $65.22$ \\
$+$ Augmentation $2\times$   & $67.70$\,\textcolor{teal}{(+2.84)} & $70.00$\,\textcolor{teal}{(+4.18)} & $67.39$\,\textcolor{teal}{(+2.17)} \\
$+$ Phase-aware modeling (\textbf{\sys}) & $\mathbf{71.39}$\,\textcolor{teal}{(+3.68)} & $\mathbf{70.33}$\,\textcolor{teal}{(+0.34)} & $\mathbf{69.83}$\,\textcolor{teal}{(+2.44)} \\
\bottomrule
\end{tabular}}
\end{table}

\begin{table}[t]
\centering
\small
\setlength{\tabcolsep}{4pt}
\caption{Augmentation-ratio sweep results without phase-aware modeling. Values are
scenario macro-F1 in percentage points.}
\label{tab:augratio}
\resizebox{0.75\columnwidth}{!}{
\begin{tabular}{lccccc}
\toprule
\textbf{Backbone} & \textbf{$0\times$} & \textbf{$1\times$} & \textbf{$2\times$} & \textbf{$4\times$} & \textbf{$8\times$} \\
\midrule
ELECTRA & $64.87$ & $66.52$ & $67.70$ & $66.50$ & $\mathbf{68.13}$ \\
RoBERTa & $65.82$ & $65.97$ & $\mathbf{70.00}$ & $69.77$ & $69.01$ \\
DeBERTa & $65.22$ & $66.45$ & $67.39$ & $\mathbf{69.08}$ & $66.46$ \\
\bottomrule
\end{tabular}}
\end{table}

\begin{table}[t]
\centering
\small
\setlength{\tabcolsep}{4pt}
\caption{Inference-time phase-aware supervision. ``Oracle'' uses the ground-truth phase label instead of the phase classifier's output.}

\label{tab:phaseinput}
\resizebox{\columnwidth}{!}{
\begin{tabular}{llcc}
\toprule
\textbf{Backbone} & \textbf{Setting} & \textbf{Scenario F1} & \textbf{Scenario Acc} \\
\midrule
\multirow{2}{*}{ELECTRA}
& Phase classifier (\sys)  & $71.39_{\pm1.60}$ & $74.96_{\pm2.04}$ \\
& Oracle  & $\mathbf{74.08}_{\pm1.57}$ & $\mathbf{79.25}_{\pm1.86}$ \\
\midrule
\multirow{2}{*}{RoBERTa}
& Phase classifier (\sys) & $70.33_{\pm1.68}$ & $73.72_{\pm1.37}$ \\
& Oracle & $\mathbf{73.30}_{\pm1.83}$ & $\mathbf{78.27}_{\pm1.37}$ \\
\midrule
\multirow{2}{*}{DeBERTa}
& Phase classifier (\sys) & $69.83_{\pm2.01}$ & $75.17_{\pm1.25}$ \\
& Oracle & $\mathbf{72.10}_{\pm2.14}$ & $\mathbf{78.84}_{\pm1.27}$ \\
\bottomrule
\end{tabular}}
\end{table}

% w/o BBB pretraining -- No much difference, better not to show
% w/o augmentation
% - different augmentation method

% \subsection{Taxonomy of Scams}
% based on the findings/stats

% - - scam category( typology): high-level theme and target, determined by the HVE
% - - scam subcategory (scenario): exact "story template"

% table, explain, example
% security company expert confirmation with real-world production.

% \subsection{Scam Scenario Classifier}
% 1. BBB classification
% - formal stuff, multi-task model architecture
% - online learning
%  improvement of each phases

% 2. Banking 

% Classification
% - human labeling and online learning on BBB
% - - 3 round of manual labeling

% use for LLM-based banking agent
% challenges:
% - stages: Unrelated, Narrative, Evidence (critical message)
% - ...
% conversation synthesize
% fine tune

\section{Discussion}

\noindent\textbf{Extendable Scenario Taxonomy.}
Our taxonomy captures the major scam scenarios observed in our dataset, but it is not exhaustive. In practice, rarer scenarios may appear infrequently, and new scenarios may emerge over time. To handle reports that are unidentifiable or too ambiguous during annotation and model training, we include an ``Other'' scenario during human annotation. Future work can study the temporal evolution of scam scenarios and update the taxonomy as new patterns emerge, which can be well supported by following our empirical study methodology.

\begin{figure}[t!]
\centering
\begin{minipage}[b]{0.45\textwidth}
\centering
\begin{tikzpicture}
\begin{axis}[
    width=\columnwidth,
    height=0.7\columnwidth,
    xbar,
    bar width=2.5pt,
    xmin=0,
    xmax=9000,
    enlarge y limits={upper=0.8, lower=0.8},
    symbolic y coords={
        Charity, Hack \& Data Breach, Sextortion, Investment \& Trading,
        Pig Butchering, Friends \& Relatives, Good Deals, Unpaid Labor,
        Tech \& Online Service, Funds{,} Grants \& Aid, Retail, Government,
        Insurance \& Warranty, Lottery \& Sweepstakes, E-commerce, Legal,
        Fake Job Offer, Financial Services
    },
    ytick=data,
    nodes near coords,
    nodes near coords style={font=\scriptsize, /pgf/number format/1000 sep=\,, inner sep=1pt},
    every node near coord/.append style={anchor=west},
    xlabel={\# of Reports},
    xlabel style={font=\scriptsize, yshift=2pt},
    tick label style={font=\scriptsize},
    y tick label style={inner sep=1pt},
    axis x line=bottom, axis y line=left,
    xmajorgrids,
    grid style={dashed, gray!25, line width=0.3pt},
    tick align=outside, tick pos=left,
    axis line style={line width=0.4pt},
    major tick length=2pt, clip=false
]
\addplot[fill=blue!60, draw=blue!70!black, line width=0.3pt] coordinates {
    (97,Charity) (150,Hack \& Data Breach) (167,Sextortion)
    (191,Investment \& Trading) (202,Pig Butchering) (289,Friends \& Relatives)
    (594,Good Deals) (771,Unpaid Labor) (1602,Tech \& Online Service)
    (1775,Funds{,} Grants \& Aid) (2283,Retail) (2398,Government)
    (2412,Insurance \& Warranty) (2537,Lottery \& Sweepstakes) (2789,E-commerce)
    (3153,Legal) (3989,Fake Job Offer) (7347,Financial Services)
};
\end{axis}
\end{tikzpicture}
\captionof{figure}{Distribution of scam scenarios.}
\label{fig:scenarioDistribution}
\end{minipage}
\hfill
\begin{minipage}[b]{0.45\textwidth}
\centering
\includegraphics[width=\textwidth]{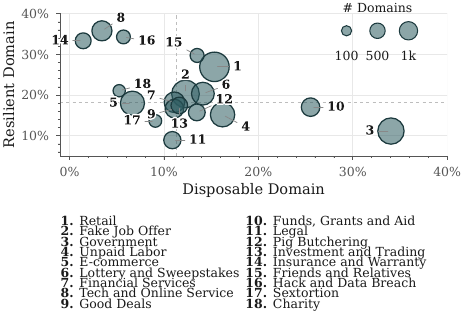}
\captionof{figure}{Resilient and disposable domain rates across scam scenarios (Finding~4).}
\label{fig:domainBubbleSub}
\end{minipage}
\end{figure}

\noindent\textbf{Unbalanced Scenario Distribution.}
As shown in \autoref{fig:scenarioDistribution}, scam scenarios follow a skewed, long-tailed distribution, which is expected in real-world scam reports but can affect detection performance. Although we use similarity-based expansion during active learning to look for more samples, some scenarios remain inherently rare in BBB Scam Tracker. Future work can mitigate this limitation by incorporating additional scam-reporting platforms~\cite{chainabuse,scamsearch} or by using synthetic data augmentation to improve coverage of low-frequency scenarios.

\noindent\textbf{Data Augmentation.}
Our two-mode LLM mutation pipeline improves robustness without requiring new labels, and the independent judge keeps label drift low. It nonetheless has inherent limits. Because mutation is label-preserving by construction, it broadens lexical and stylistic coverage within existing scenarios but cannot introduce genuinely new scenarios or improve coverage of the long tail. Our ablation shows diminishing returns beyond a $2\times$ ratio (\autoref{tab:augratio}). Future work could explore controllable or persona-conditioned generation, augmentation targeted at rare scenarios, and mixing in additional real conversations under the same privacy constraints.

\noindent\textbf{Conversation Context Window.}
Our detector inherits the $512$-token context limit of base-size transformer encoders, and we cap inputs at $\texttt{max\_turns}=64$ with a $512$-token prefix budget. In our corpus this is usually sufficient (mean $12.9$ turns and $\approx\!350$ sub-word tokens per conversation), but the longest conversations are truncated, which drops earlier context that may carry diagnostic evidence, and real customer-service conversations can be longer still. Several directions could relax this constraint: (i) long-context encoders that widen the window directly;  (ii) hierarchical or streaming encoding that encodes each turn once and aggregates across turns, which both avoids re-truncating long prefixes and removes the cost of re-encoding the entire prefix at every turn during live, turn-by-turn deployment; and (iii) compress or summarize historical turns by following research in LLM agent memory~\cite{amem}. 

% \noident\textbf{Binary vs Multiclass Classification.} 
% we do not conversation is benign or scam-related. assume the scam already took place and detect which scenario it is . this information is ued to assist banker's responsibiltiy to  by checking whether the payment is fraudant. 

\section{Related Work}
\label{sec:literature}

\noindent\textbf{Empirical Studies of Online Scams.}
Most prior efforts focus on specific forms of online scams. Park et al.~\cite{park2014scambaiter} studied targeted Nigerian scams on Craigslist using honeypot advertisements and automated scammer interactions. Miramirkhani et al.~\cite{miramirkhani2017dial} and Acharya et al.~\cite{acharya2024conning} conducted large-scale studies of technical support scams, revealing how scammers combine deceptive webpages, malvertising, phone calls, remote-access tools, and payment channels to monetize victims. Acharya and Holz~\cite{acharya2024explorative} examined pig-butchering scams, highlighting their long-term grooming process and severe financial harm. Kharraz et al.~\cite{kharraz2018surveylance} and Ma et al.~\cite{ma2024careful} studied web- and mobile-ad-driven scam campaigns, showing that such campaigns not only monetize user engagement but also expose users to broader security threats, including malware and identity theft. These studies provide valuable insights into individual scam types, but lack a comprehensive measurement of the broader scam ecosystem.

% \noindent\textbf{Social Engineering in Scams.}

\noindent\textbf{Early Classification in Harmful Conversations.}
 Prior work has studied early classification as a sequential
decision problem, in which a model must balance prediction accuracy with the cost
of waiting for more evidence~\cite{dulacarnold2011sequential,burdisso2019ss3,burdisso2020tss3}. Dialogue classification work further shows that
multi-turn context is essential for understanding intent, risk, and user state,
because critical evidence may be distributed across turns rather than appearing
in a single utterance~\cite{serban2016hierarchical,raheja2019dialogue,
qu2019intent}.
Similar ideas have been applied to harmful
conversation detection, such as identifying risky online interactions or abusive
dialogues from partial conversation prefixes~\cite{an2025revisiting,chehbouni2025enhancing}.
Building on these insights, our work studies early scam scenario detection in customer-service conversations, where the defender must classify a scam scenario from a progressively revealed customer narrative.

\noindent\textbf{Conversation-based Scam Detection.}
The most relevant research direction to our setting is conversation-based scam detection. 
Traditional methods detect vishing or fraud from conversation transcripts using handcrafted signals, such as clustering scam signatures derived from speech acts~\cite{derakhshan2021detecting} or applying decision trees over linguistic features~\cite{bajaj2019fraud}. 
Recent efforts further explore LLMs for real-time scam detection from ongoing conversations~\cite{shen2025warned,sun2026prescam}. 
Our setting differs because the defender observes a second-hand customer narrative elicited by a bank representative, rather than the original attacker-victim conversation. 

\section{Conclusion}
In this paper, we conduct a large-scale study of scam scenarios using 102,054 real-world incident reports and construct a hierarchical taxonomy of 18 scenarios grouped into 6 high-level tactics based on associated PTs. Our analysis reveals that scammers reuse PTs, scenarios, and infrastructure across diverse contexts to scale their operations. Building on these insights, we develop a conversation-aware scam scenario detection method for customer-service conversations that enables early detection from partial and evolving dialogues. Experiments on 1,115 conversations synthesized from real customer-service calls demonstrate that our approach effectively identifies scam scenarios in practical settings and that scenario-level knowledge provides a practical and interpretable foundation for both analysis and defense.

% \input{acknowledgement}

%-------------------------------------------------------------------------------
% \bibliographystyle{ACM-Reference-Format}
\bibliographystyle{IEEEtran}
\bibliography{./bibliography/refs}
% \newpage
\appendices

\section{Human Review Protocol}
\label{app:humanreviewprotocol}

Each annotation instance corresponds to one scam incident report. Annotators
are asked to assign one dominant scenario label from our taxonomy. When a
report contains multiple scam elements, annotators label the scenario that
best captures the scammer's primary pretext, victim-facing narrative, and
requested action. If no scenario applies, annotators assign the label
``Other.'' If the report does not contain enough information to make a
confident decision, annotators mark it as ``need review'' for later
adjudication.

\noindent\textbf{Guidelines and Decision Rules.}
Annotators follow a two-step decision process. First, they identify the
high-level tactic by examining the prominent PTs
used in scammer's pretext. Second, within the identified tactic, annotators examine the report content to select the most appropriate scenario.

\noindent\textbf{Agreement and Adjudication.}
To assess the reliability of the human annotation process, we measure
inter-annotator agreement on the double-labeled reports using Cohen's
$\kappa$, which accounts for agreement that may occur by chance. The two
primary annotators achieved substantial agreement
($\kappa=0.75$), indicating that the scenario taxonomy and annotation
guideline were consistently understood and applied. Cases with disagreement
were further reviewed by a third annotator. If two annotators agreed after
adjudication, the majority label was used as the final label; otherwise, the
case was discussed jointly until consensus was reached. This process helps
ensure that the final scenario labels are both reliable and grounded in the
report narratives. %\autoref{fig:humanAnnotationInterface} shows the annotation interface
%of an annotation page and a review page.

% \begin{figure*}[h]
%     \centering

%     \begin{subfigure}{0.48\linewidth}
%         \centering
%         \includegraphics[width=\linewidth]{figs/CodingInterface.png}
%         \caption{Annotation page}
%         \label{fig:annotation-page}
%     \end{subfigure}
%     \hfill
%     \begin{subfigure}{0.48\linewidth}
%         \centering
%         \includegraphics[width=\linewidth]{figs/ReviewInterface.png}
%         \caption{Review page}
%         \label{fig:review-page}
%     \end{subfigure}

%     \caption{Examples of the human annotation interface.}
%     \label{fig:humanAnnotationInterface}
% \end{figure*}
\section{Additional Figure and Tables}
\label{appendix:pt}

% \begin{table}[h!]
% \centering
% \caption{Phase-state features.}
% \label{tab:phasefeatures}
% \scriptsize
% \setlength{\tabcolsep}{1pt}
% \renewcommand{\arraystretch}{0.95}
% % \resizebox{\columnwidth}{!}{
% \begin{tabular}{@{}rll@{}}
% \toprule
% \# & Feature & Description \\
% \midrule
% 1  & \texttt{cur\_p\_unrelated}   & Current turn is unrelated. \\
% 2  & \texttt{cur\_p\_narrative}   & Current turn provides narrative context. \\
% 3  & \texttt{cur\_p\_evidence}    & Current turn contains scam evidence. \\
% 4  & \texttt{pre\_mean\_p\_narr}  & Narrative share in the prefix. \\
% 5  & \texttt{pre\_mean\_p\_evid}  & Evidence share in the prefix. \\
% 6  & \texttt{pre\_max\_p\_evid}   & Whether evidence has appeared. \\
% 7  & \texttt{related\_frac}       & Share of scam-related turns. \\
% 8  & \texttt{evidence\_frac}      & Share of evidence turns. \\
% 9  & \texttt{first\_evid\_norm}   & Normalized first-evidence position. \\
% 10 & \texttt{since\_evid\_norm}   & Normalized distance from first evidence. \\
% 11 & \texttt{narr2evid\_frac}     & Rate of narrative-to-evidence transitions. \\
% \midrule
% 12 & \texttt{evid\_offset}        & Offset $d_t=t-t_e$ to evidence onset. \\
% \bottomrule
% \end{tabular}
% % }
% \end{table}

\begin{table}[h!]
\centering
\caption{The two phase-aware signals constructed by the Phase Info Builder.}
\label{tab:phasefeatures}
\scriptsize
\setlength{\tabcolsep}{1.8pt}
\renewcommand{\arraystretch}{0.78}
\begin{tabular}{@{}rlp{0.6\linewidth}@{}}
\toprule
\# & Feature & Description \\
\midrule
1  & \texttt{cur\_p\_unrelated} & Indicator that the current turn is \textsc{Unrelated}. \\
2  & \texttt{cur\_p\_narrative} & Indicator that the current turn is \textsc{Narrative}. \\
3  & \texttt{cur\_p\_evidence} & Indicator that the current turn is \textsc{Scam\_Evidence}. \\
4  & \texttt{pre\_mean\_p\_narr} & Running mean of the narrative channel over the prefix, i.e., the share of \textsc{Narrative} turns so far. \\
5  & \texttt{pre\_mean\_p\_evid} & Running mean of the evidence channel over the prefix. \\
6  & \texttt{pre\_max\_p\_evid} & Running maximum of the evidence channel, i.e., whether an evidence turn has appeared in the prefix. \\
7  & \texttt{related\_frac} & Fraction of prefix turns labeled scam-related, i.e., \textsc{Narrative} or \textsc{Scam\_Evidence}. \\
8  & \texttt{evidence\_frac} & Fraction of prefix turns labeled \textsc{Scam\_Evidence}. \\
9  & \texttt{first\_evid\_norm} & Normalized position of the first evidence turn within the prefix; $1$ if no evidence turn has appeared yet. \\
10 & \texttt{since\_evid\_norm} & Normalized number of turns elapsed since the first evidence turn; $0$ before evidence appears. \\
11 & \texttt{narr2evid\_frac} & Fraction of adjacent prefix turn pairs that switch from \textsc{Narrative} to \textsc{Scam\_Evidence}, capturing escalation into evidence. \\
\midrule
12 & \texttt{evid\_offset} & Signed offset $d_t=t-t_e$ to the \textsc{Scam\_Evidence} onset turn. Mapped to the hidden space by the learnable embedding $e_{\mathrm{off}}(d_t)$. \\
\bottomrule
\end{tabular}
\end{table}

\begin{table}[t!]
\centering
\scriptsize
\setlength{\tabcolsep}{3pt}
\renewcommand{\arraystretch}{1.0}
\caption{Customer-service conversation dataset statistics. \#All is the total conversation count. Turns is
the mean number of turns per conversation. $t_e$ the mean annotated onset turn of \textsc{Scam_Evidence}.}
\label{tab:convdataset}
\begin{tabular}{@{}lrrr@{}}
\toprule
\textbf{Tactic / Scenario} & \textbf{\#All} & \textbf{Turns} & \textbf{$t_e$} \\
\midrule
\textbf{Authority \& Compliance} \\
\quad Government               & 109 & 10.7 & 3.3 \\
\quad Legal                    &  65 & 11.8 & 3.2 \\
\quad Other Authority          &  36 & 13.5 & 4.2 \\
\textbf{Windfall} \\
\quad Good Deals               &  71 & 11.7 & 4.2 \\
\quad Investment \& Trading    &  45 & 14.4 & 3.8 \\
\quad Fund, Grants \& Aid      &  23 & 11.8 & 2.7 \\
\quad Lottery \& Sweepstakes   &  20 & 11.1 & 3.0 \\
\textbf{Consumer \& Services} \\
\quad Financial Services       &  70 & 11.9 & 3.0 \\
\quad Tech \& Online Service   &  58 & 13.4 & 3.7 \\
\quad E-commerce               &  47 & 12.1 & 3.7 \\
\quad Retail                   &  43 & 12.3 & 3.8 \\
\quad Insurance \& Warranty    &  24 & 10.2 & 3.6 \\
\textbf{Employment} \\
\quad Fake Job Offer           & 135 & 12.7 & 3.1 \\
\quad Unpaid Labor             &  87 & 14.9 & 4.8 \\
\textbf{Relationship \& Trust} \\
\quad Pig Butchering           &  45 & 16.6 & 5.3 \\
\quad Friend \& Relative       &  41 & 12.8 & 3.9 \\
\quad Charity                  &  29 & 12.2 & 4.4 \\
\quad Other Relationship       &  51 & 14.4 & 3.6 \\
\textbf{Extortion} \\
\quad Sextortion               &  68 & 14.1 & 3.9 \\
\quad Hack \& Data Breach      &  48 & 15.0 & 4.1 \\
\midrule
\textbf{Total} & \textbf{1{,}115} & \textbf{12.9} & \textbf{3.8} \\
\bottomrule
\end{tabular}
\end{table}

\begin{table}[t!]
\centering
\scriptsize
\setlength{\tabcolsep}{2.5pt}
\renewcommand{\arraystretch}{0.92}
\caption{F1 score of scam incident report classification (\%).}
\label{tab:scenarioSubcategoryResults}
\begin{tabular}{@{}l@{\hspace{4pt}}rccc@{}}
\toprule
\textbf{Tactic / Scenario} & \textbf{Sup.} & \textbf{RoB.} & \textbf{ELEC.} & \textbf{DeB.} \\
\midrule
\textbf{Authority \& Compliance} & 61 & \textbf{91.63} & \textbf{91.16} & \textbf{91.41} \\
\quad Government               & 29 & 91.92 & 88.57 & 90.83 \\
\quad Legal                    & 32 & 89.47 & 87.88 & 91.64 \\
\textbf{Consumer \& Services} & 137 & \textbf{88.10} & \textbf{89.50} & \textbf{89.24} \\
\quad E-commerce               & 24 & 86.65 & 86.29 & 85.73 \\
\quad Financial Services       & 34 & 72.40 & 67.74 & 75.07 \\
\quad Insurance \& Warranty    & 27 & 91.14 & 92.07 & 90.13 \\
\quad Retail                   & 26 & 74.60 & 69.19 & 68.29 \\
\quad Tech \& Online Service   & 26 & 81.33 & 84.24 & 82.48 \\
\textbf{Employment} & 90 & \textbf{96.92} & \textbf{95.50} & \textbf{95.80} \\
\quad Fake Job Offer           & 50 & 89.41 & 87.43 & 88.68 \\
\quad Unpaid Labor             & 40 & 84.48 & 84.50 & 83.69 \\
\textbf{Extortion} & 24 & \textbf{91.74} & \textbf{94.84} & \textbf{93.81} \\
\quad Hack \& Data Breach      &  9 & 23.04 & 45.16 & 70.71 \\
\quad Sextortion               & 15 & 87.94 & 87.77 & 94.84 \\
\textbf{Relationship \& Trust} & 42 & \textbf{87.50} & \textbf{88.90} & \textbf{91.06} \\
\quad Charity                  & 14 & 74.66 & 83.92 & 92.18 \\
\quad Friend and Relative      & 14 & 71.99 & 74.33 & 84.49 \\
\quad Pig Butchering           & 14 & 77.24 & 79.56 & 85.75 \\
\textbf{Windfall} & 107 & \textbf{90.05} & \textbf{92.15} & \textbf{91.01} \\
\quad Fund, Grants and Aid     & 26 & 87.70 & 91.63 & 88.93 \\
\quad Good Deals               & 20 & 81.32 & 83.12 & 87.58 \\
\quad Investment and Trading   & 16 & 74.10 & 79.54 & 86.24 \\
\quad Lottery and Sweepstakes  & 45 & 86.36 & 88.34 & 86.90 \\
\midrule
\textit{Overall Tactic}    & 461 & 90.99 & 92.01 & 92.05 \\
\textit{Overall Scenario} & 461 & 79.21 & 81.18 & 85.23 \\
\bottomrule
\end{tabular}
\end{table}

\begin{table*}[t!]
\centering
\scriptsize
\setlength{\tabcolsep}{1pt}
\renewcommand{\arraystretch}{1.1}
\caption{Definition of the psychological techniques used in this paper.}
\begin{tabularx}{\linewidth}{
    >{\raggedright\arraybackslash}p{0.20\linewidth}
    >{\raggedright\arraybackslash}X
}
\toprule
\textbf{Psychological Technique} & \textbf{Description} \\
\midrule
Authority & Leveraging the tendency of people to obey authority figures. Scammers impersonate roles such as government officials, executives, police officers, lawyers, doctors, or religious leaders to gain compliance. \\
Phantom Riches & Exploiting desire and reward-seeking by presenting promises of large gains, easy money, or unusually attractive opportunities that can override rational judgment. \\
Fear and Intimidation & Triggering fear, panic, or anxiety to pressure victims into immediate compliance and suppress careful reasoning. \\
Liking & Persuading victims by appearing friendly, charming, supportive, or flattering, thereby increasing rapport and reducing suspicion. \\
Urgency and Scarcity & Creating a sense of time pressure or limited availability so that victims act quickly without sufficient verification or reflection. \\
Credibility & Fabricating a credible story, scenario, or identity to establish legitimacy and gain the victim's trust. \\
Evoking Social Norms & Exploiting socially desirable norms such as kindness, helpfulness, reciprocity, and obligation, e.g., making victims feel they should return a favor or cooperate. \\
Consistency & Exploiting the tendency of people to behave consistently with their prior commitments, statements, or actions. \\
Social Proof & Influencing victims by referencing the behavior, approval, or participation of others, suggesting that the scam is legitimate because many people are doing it. \\
\bottomrule
\end{tabularx}
\label{tab:ptdefinitions}
\end{table*}

% \section{Details of Empirical Study}
% \label{app:empiricaldetails}

\begin{table*}[t!]
\centering
\scriptsize
\setlength{\tabcolsep}{4pt}
\renewcommand{\arraystretch}{1.12}
\caption{Raw outputs of the BERTopic model. The raw topics provide candidate clusters for scenario construction rather than final scenario labels. Some topics exhibit semantic overlap, such as loan-related Topics~3 and~6, and job-related Topics~5,~14, and~16. This motivates the subsequent manual inspection, merging, and refinement steps used to construct the final scam scenario taxonomy.}
\begin{tabularx}{\textwidth}{r r p{0.38\textwidth} X}
\toprule
\textbf{Topic} & \textbf{Reports} & \textbf{Top-10 keywords} & \textbf{LLM summary based on top-50 keywords} \\
\midrule
0  & 8,524 & card, bank, check, told, flight, credit, credit card, gift, cards, won
& Mixed payment and consumer scams involving credit cards, bank checks, prizes, travel reservations, gift cards, or cancellation-related charges. \\

1  & 5,978 & warranty, letter, coverage, home, home warranty, notice, property, vehicle, mortgage, final
& Warranty-renewal or coverage notice scams that pressure victims through mail about home, vehicle, mortgage, or property protection plans. \\

2  & 5,407 & leave, days, earn, merchants, days paid, paid, paid annual, parttime, annual leave, paid annual leave
& Remote part-time job scams promising easy earnings, flexible schedules, paid leave, and short daily work for merchant visibility or similar tasks. \\

3  & 4,408 & loan, calls, numbers, different, calling, approved, times, different numbers, applied, list
& Loan-related robocall or telemarketing scams that repeatedly contact victims from different numbers with claims of loan approval. \\

4  & 4,395 & court, case, office, documents, legal, place, calling, served, voicemail, filing
& Legal threat scams that claim victims have court cases, legal documents, filings, or pending service of process to induce fear and response. \\

5  & 3,880 & packages, job, interview, package, shipping, position, month, offer, linkedin, check
& Fake job and package-reshipping scams where victims are recruited for shipping-related positions and asked to receive or forward packages. \\

6  & 2,495 & press, loan, approval, underwriting, connect, opt, personal loan, just need, quick, longer
& Personal loan approval scams using scripted messages that ask victims to confirm information, connect with agents, or proceed with funding. \\

7  & 2,443 & refund, return, product, ordered, china, item, dress, items, shoes, order
& Online shopping scams involving low-quality, counterfeit, or misrepresented products, often followed by failed refund or return requests. \\

8  & 2,388 & withdraw, trading, funds, crypto, investment, platform, withdrawal, deposit, invest, wallet
& Crypto, trading, and investment scams where victims are shown profits but are blocked from withdrawing funds without additional deposits or fees. \\

9  & 1,645 & invoice, norton, subscription, mcafee, geek, renewal, payment, geek squad, squad, consulting
& Fake subscription renewal and tech-support invoice scams impersonating Norton, McAfee, Geek Squad, or similar service providers. \\

10 & 1,358 & facebook, site, instagram, fake, people, page, scammer, fb, real, using
& Social media impersonation scams using fake Facebook or Instagram profiles, pages, ads, photos, or hacked accounts to lure victims. \\

11 & 1,296 & medicare, insurance, medical, health, health insurance, doctor, patient, knee, hospital, plan
& Medicare, health insurance, and medical billing scams involving insurance plans, doctors, medical supplies, labs, or patient information. \\

12 & 1,287 & door, garage, car, driveway, garage door, parking, repair, told, contractor, price
& Local service scams involving garage doors, car repairs, towing, contractors, locks, or home repair work with inflated prices or incomplete jobs. \\

13 & 1,260 & form, beneficial, ownership, letter, filing, annual, records service, mandatory, reporting, ein
& Business compliance filing scams that send official-looking letters about beneficial ownership, annual filings, records services, or penalties. \\

14 & 1,011 & interview, position, teams, microsoft teams, microsoft, hiring, manager, entry, data entry, data
& Fake hiring scams for data-entry or remote positions, often conducted through Microsoft Teams, Signal, or fake HR/hiring manager personas. \\

15 & 938 & paypal, invoice, transaction, paypal account, usd, received email, invoice number, order, purchase, bitcoin
& PayPal invoice scams that send fake transaction or purchase notices, often involving unauthorized charges, Bitcoin, or seller payment claims. \\

16 & 904 & trial period, trial, period, hr, global streaming, wbd global, wbd, streaming, 3day paid, 3day
& Fake streaming or media-company job scams advertising paid trial periods, daily tasks, promotions, and customer service roles. \\

17 & 832 & passport, government, renewal, site, application, renew, form, government website, global, personal information
& Fake government service websites for passport renewal or applications that collect personal information, fees, and credit card details. \\

18 & 789 & toll, avoid, unpaid, late, unpaid toll, balance, settle, link, late fees, outstanding
& Toll-payment phishing scams that send messages about unpaid tolls, late fees, or outstanding balances and direct victims to malicious links. \\

19 & 771 & tax, filings, tax debt, end month, settled, taxes, wage, wage garnishment, garnishment, overdue
& Tax debt relief scams claiming to reduce overdue taxes, stop wage garnishment, or settle tax filings through relief programs. \\
\bottomrule
\end{tabularx}

\label{tab:bertopictopicssummary}
\end{table*}

% \newpage
% \section{LLM Prompts}
% \label{app:llmprompts}

% =============================================================================
% Figure: Prompt template for LLM-based topic summarization
% Reuses the `systemprompt` / `userprompt` tcolorbox styles and the \ph{...}
% and \jkey{...} commands defined alongside the scenario-annotation prompt
% figure. If this is the first prompt figure in your paper, also include the
% preamble block from prompt_figure.tex.
% =============================================================================
\definecolor{promptSysBg}{RGB}{242, 247, 252}
\definecolor{promptSysBorder}{RGB}{52, 103, 158}
\definecolor{promptUserBg}{RGB}{252, 247, 240}
\definecolor{promptUserBorder}{RGB}{178, 110, 38}
\definecolor{promptPlaceholder}{RGB}{160, 40, 100}
\definecolor{promptKey}{RGB}{40, 90, 150}

% --- Reusable styles (place once in preamble) ---
\newtcolorbox{systemprompt}{
    colback=promptSysBg,
    colframe=promptSysBorder,
    boxrule=0.5pt,
    arc=2pt, outer arc=2pt,
    left=6pt, right=6pt, top=4pt, bottom=4pt,
    fonttitle=\bfseries\small,
    title={\textsc{System Prompt}},
    coltitle=white,
    colbacktitle=promptSysBorder,
    % attach boxed title to top left={yshift=-2pt, xshift=4pt},
    % boxed title style={
    %     colback=promptSysBorder,
    %     boxrule=0pt, arc=1pt,
    %     left=4pt, right=4pt, top=1pt, bottom=1pt
    % },
    top=10pt
}

\newtcolorbox{userprompt}{
    % enhanced, breakable,
    colback=promptUserBg,
    colframe=promptUserBorder,
    boxrule=0.5pt,
    arc=2pt, outer arc=2pt,
    left=6pt, right=6pt, top=4pt, bottom=4pt,
    fonttitle=\bfseries\small,
    title={\textsc{User Prompt}},
    coltitle=white,
    colbacktitle=promptUserBorder,
    % attach boxed title to top left={yshift=-2pt, xshift=4pt},
    % boxed title style={
    %     colback=promptUserBorder,
    %     boxrule=0pt, arc=1pt,
    %     left=4pt, right=4pt, top=1pt, bottom=1pt
    % },
    top=10pt
}

\newtcolorbox{normalprompt}{
    % enhanced, breakable,
    colback=promptUserBg,
    colframe=promptUserBorder,
    boxrule=0.5pt,
    arc=2pt, outer arc=2pt,
    left=6pt, right=6pt, top=4pt, bottom=4pt,
    fonttitle=\bfseries\small,
    title={\textsc{Prompt}},
    coltitle=white,
    colbacktitle=promptUserBorder,
    % attach boxed title to top left={yshift=-2pt, xshift=4pt},
    % boxed title style={
    %     colback=promptUserBorder,
    %     boxrule=0pt, arc=1pt,
    %     left=4pt, right=4pt, top=1pt, bottom=1pt
    % },
    top=10pt
}

% Placeholder command for {variable} slots
\newcommand{\ph}[1]{\textcolor{promptPlaceholder}{\textit{\{#1\}}}}
% Key command for JSON keys
\newcommand{\jkey}[1]{\textcolor{promptKey}{\texttt{"#1"}}}

% \begin{table}[h]
% \centering
% \caption{Training hyperparameters for the phase-aware turn-level classifier.}
% \label{tab:hyperparams}
% \footnotesize
% \begin{tabular}{@{}lll@{}}
% \toprule
% Group & Hyperparameter & Value \\
% \midrule
% \multirow{6}{*}{Optimization}
%  & optimizer & AdamW \\
%  & weight decay & $0.01$ \\
%  & encoder LR & $1\times10^{-5}$ \\
%  & head / injection LR & $8\times10^{-5}$ \\
%  & batch size & $4$ \\
%  & epochs & $10$ \\
% \midrule
% \multirow{3}{*}{Sequence}
%  & \texttt{max\_turns} & $64$ \\
%  & \texttt{max\_turn\_length} & $128$ \\
%  & \texttt{max\_prefix\_length} & $512$ \\
% \midrule
% \multirow{2}{*}{Encoder}
%  & hidden size & $768$ \\
%  & dropout & $0.1$ \\
% \midrule
% \multirow{3}{*}{Phase injection}
%  & phase-state dim & $11$ \\
%  & evidence offset range & $\pm 12$ \\
%  & phase input & one-hot \\
% \midrule
% \multirow{4}{*}{Loss}
%  & \texttt{legit\_turn\_weight} & $1.0$ \\
%  & \texttt{scam\_turn\_weight} & $1.5$ \\
%  & \texttt{onset\_turn\_weight} & $2.5$ \\
%  & \texttt{class\_balance\_power} & $0.5$ \\
% \midrule
% \multirow{1}{*}{Augmentation}
%  & ratio & $2\times$ source pool \\
% \bottomrule
% \end{tabular}
% \end{table}

\begin{figure}[t!]
\centering
\scriptsize
\begin{systemprompt}
You are an expert at analyzing scam incident reports. A \textbf{scam scenario} is the recurring operational scheme that captures the core strategy used by the scammer, including the claimed pretext, the victim-facing narrative, and the requested action or payment method. Given a scam incident report, your task is to identify the scam scenario that best describes the underlying scheme. The complete taxonomy of considered scenarios is provided below.

% \medskip
\ph{scenario taxonomy}
\end{systemprompt}

% \vspace{4pt}

\begin{userprompt}
\textbf{Scam Incident Report:} \ph{description}

% \medskip
\textbf{Analysis Steps:}
\begin{enumerate}[leftmargin=*, itemsep=1pt, topsep=2pt]
    \item Identify the dominant psychological technique that makes this scam succeed, and use it to determine the underlying \emph{tactic}.
    \item Based on the victim-facing narrative, select the most appropriate \emph{scenario} from the taxonomy.
\end{enumerate}

\textbf{Output Requirements:} Provide a concise summary of your reasoning, and return the result strictly as a JSON object:
{\ttfamily\tiny \{ \jkey{tactic}: \ph{tactic\_name} \textbar{} \texttt{"Other"}, \jkey{scenario}: \ph{scenario\_name} \textbar{} \texttt{"Other"}, \jkey{summary}: \ph{brief classification rationale}, \jkey{reference}: \ph{scam-relevant content} \}}

\medskip
\ph{few-shot examples}
\end{userprompt}

\caption{Prompt template used by the LLM annotator to produce draft scenario labels for scam incident reports.}
\label{fig:annotationPrompt}
\end{figure}

% \begin{figure}[t!]
% \centering
% \footnotesize
% \begin{tcolorbox}[
%   colback=gray!3, colframe=black!55, boxrule=0.4pt, arc=2pt,
%   left=5pt, right=5pt, top=4pt, bottom=4pt,
%   title=\small\textbf{\textcolor{white}{\nolinkurl{https://www.bbb.org/scamtracker/lookupscam/1259631}}},
%   fonttitle=\bfseries, coltitle=white, colbacktitle=black!55
% ]

% \textbf{Description.} They told us we would have our power cut off at
% 5\,pm if we did not give them money. They wouldn't take it through the
% website of Con-Edison---we would have to go to a Walgreens or CVS.
% They told us we missed all of the text and email warnings of
% disconnect and it would be cut off until Monday. They told us to call
% 800-305-6276 ext.\ 1. My case number was *******.

% \vspace{3pt}\hrule\vspace{3pt}

% {\setlength{\tabcolsep}{3pt}\renewcommand{\arraystretch}{1.05}%
% \begin{tabular}{@{}p{0.47\columnwidth}p{0.47\columnwidth}@{}}
% \textbf{Scam Type:} Utility      & \textbf{Scam ID:} 1259631        \\
% \textbf{Scammer:} NY, (718) 874-9444 & \textbf{Victim Loc.:} NY 11205  \\
% \textbf{Date:} April 17, 2026    & \textbf{Email / URL:} unknown    \\
% \end{tabular}}
% \end{tcolorbox}
% \vspace{-4pt}
% \caption{A real scam incident report for the utility impersonation-gift card scam in \autoref{fig:exampleConversation}.}
% \label{fig:exampleUtilityReport}
% \end{figure}

\begin{figure}[t!]
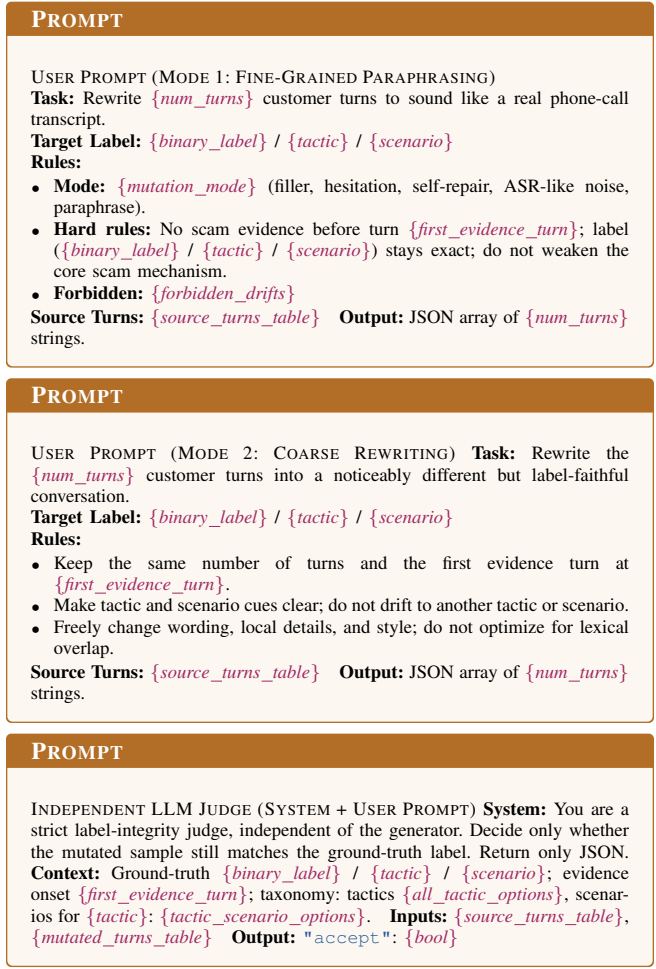

\centering
\scriptsize

% \begin{systemprompt}\textsc{System Prompt (Shared by Both Generator Modes)}
% You are a conversation mutation engine for ML training-data augmentation. Follow the task instructions exactly; no commentary. Return only the requested JSON.
% \end{systemprompt}

% \vspace{3pt}

% \begin{minipage}[t]{0.485\linewidth}
\begin{normalprompt}\textsc{User Prompt (Mode 1: Fine-Grained Paraphrasing)}

\textbf{Task:} Rewrite \ph{num\_turns} customer turns to sound like a real phone-call transcript.

\textbf{Target Label:} \ph{binary\_label} / \ph{tactic} / \ph{scenario}

\textbf{Rules:}
\begin{itemize}[leftmargin=*, itemsep=0pt, topsep=1pt, parsep=0pt]
    \item \textbf{Mode:} \ph{mutation\_mode} (filler, hesitation, self-repair, ASR-like noise, paraphrase).
    \item \textbf{Hard rules:} No scam evidence before turn \ph{first\_evidence\_turn}; label (\ph{binary\_label} / \ph{tactic} / \ph{scenario}) stays exact; do not weaken the core scam mechanism.
    \item \textbf{Forbidden:} \ph{forbidden\_drifts}
\end{itemize}

\textbf{Source Turns:} \ph{source\_turns\_table}\quad
\textbf{Output:} JSON array of \ph{num\_turns} strings.

% \textbf{Self-Check:} \jkey{evidence\_boundary\_preserved}, \jkey{scenario\_preserved}: \ph{bool}; \jkey{risk\_note}: \ph{string}
\end{normalprompt}
% \end{minipage}%
% \hfill
% \begin{minipage}[t]{0.485\linewidth}
\begin{normalprompt}\textsc{User Prompt (Mode 2: Coarse Rewriting)}
\textbf{Task:} Rewrite the \ph{num\_turns} customer turns into a noticeably different but label-faithful conversation.

\textbf{Target Label:} \ph{binary\_label} / \ph{tactic} / \ph{scenario}

\textbf{Rules:}
\begin{itemize}[leftmargin=*, itemsep=0pt, topsep=1pt, parsep=0pt]
    \item Keep the same number of turns and the first evidence turn at \ph{first\_evidence\_turn}.
    \item Make tactic and scenario cues clear; do not drift to another tactic or scenario.
    \item Freely change wording, local details, and style; do not optimize for lexical overlap.
\end{itemize}

\textbf{Source Turns:} \ph{source\_turns\_table}\quad
\textbf{Output:} JSON array of \ph{num\_turns} strings.
\end{normalprompt}
% \end{minipage}

\begin{normalprompt}\textsc{Independent LLM Judge (System + User Prompt)}
\textbf{System:} You are a strict label-integrity judge, independent of the generator. Decide only whether the mutated sample still matches the ground-truth label. Return only JSON.\quad
\textbf{Context:} Ground-truth \ph{binary\_label} / \ph{tactic} / \ph{scenario}; evidence onset \ph{first\_evidence\_turn}; taxonomy: tactics \ph{all\_tactic\_options}, scenarios for \ph{tactic}: \ph{tactic\_scenario\_options}.\quad
\textbf{Inputs:} \ph{source\_turns\_table}, \ph{mutated\_turns\_table}\quad
\textbf{Output:} \jkey{accept}: \ph{bool}
\end{normalprompt}

\caption{Prompt templates for LLM-based data augmentation.}
\label{fig:augmentationPrompts}
\end{figure}

\begin{figure}[t!]
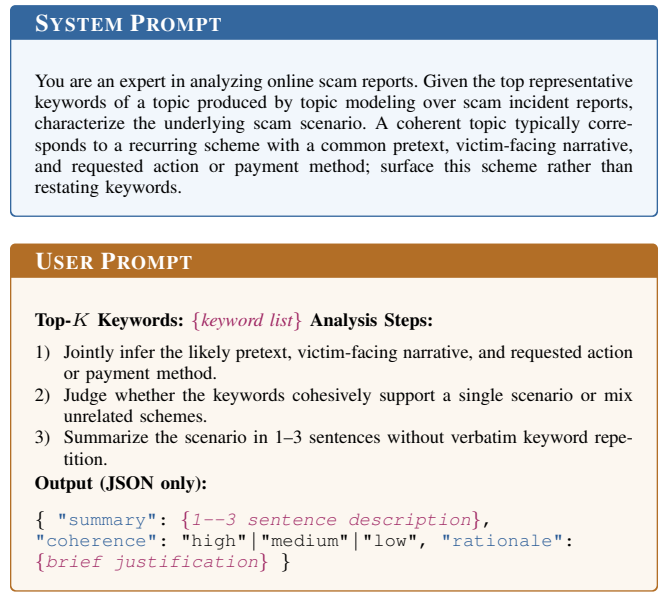

\centering
\scriptsize
\begin{systemprompt}
You are an expert in analyzing online scam reports. Given the top representative keywords of a topic produced by topic modeling over scam incident reports, characterize the underlying scam scenario. A coherent topic typically corresponds to a recurring scheme with a common pretext, victim-facing narrative, and requested action or payment method; surface this scheme rather than restating keywords.
\end{systemprompt}
\vspace{2pt}
\begin{userprompt}
\textbf{Top-$K$ Keywords:} \ph{keyword list}
\smallskip
\textbf{Analysis Steps:}
\begin{enumerate}[leftmargin=*, itemsep=0pt, topsep=1pt]
    \item Jointly infer the likely pretext, victim-facing narrative, and requested action or payment method.
    \item Judge whether the keywords cohesively support a single scenario or mix unrelated schemes.
    \item Summarize the scenario in 1--3 sentences without verbatim keyword repetition.
\end{enumerate} \textbf{Output (JSON only):}
\begin{flushleft}
\ttfamily
\{ \jkey{summary}: \ph{1--3 sentence description}, \jkey{coherence}: \texttt{"high"}\,\textbar\,\texttt{"medium"}\,\textbar\,\texttt{"low"}, \jkey{rationale}: \ph{brief justification} \}
\end{flushleft}
\end{userprompt}
\caption{Prompt template for generating scenario summaries for each BERTopic cluster.}
\label{fig:topicSummaryPrompt}
\end{figure}

%%%%%%%%%%%%%%%%%%%%%%%%%%%%%%%%%%%%%%%%%%%%%%%%%%%%%%%%%%%%%%%%%%%%%%%%%%%%%%%%
\end{document}